%% file: main.tex
\documentclass[10pt,pre,twocolumn,tightenlines,longbibliography,notitlepage, nofootinbib,superscriptaddress]{revtex4-2}
\pdfoutput=1
 
\usepackage{xcolor}
\usepackage{verbatim}
\usepackage{graphicx}
\usepackage{natbib}
\usepackage{siunitx}
\usepackage[textsize=tiny]{todonotes}

\newcommand{\kT}{k$_{\mathrm{B}}T$}

\begin{document}

\title{Economical routes to size-specific assembly of self-closing structures}

\author{Thomas E. Videb\ae k}
\email{videbaek@brandeis.edu}

\author{Daichi Hayakawa}
\affiliation{Martin A. Fisher School of Physics, Brandeis University, Waltham, MA, 02453, USA}
\author{Gregory M. Grason}
\affiliation{Department of Polymer Science and Engineering, University of Massachusetts, Amherst, Massachusetts 01003, USA}
\author{Michael F. Hagan}
\affiliation{Martin A. Fisher School of Physics, Brandeis University, Waltham, MA, 02453, USA}
\author{Seth Fraden}
\affiliation{Martin A. Fisher School of Physics, Brandeis University, Waltham, MA, 02453, USA}
\author{W. Benjamin Rogers}
\email{wrogers@brandeis.edu}
\affiliation{Martin A. Fisher School of Physics, Brandeis University, Waltham, MA, 02453, USA}

\begin{abstract}
Self-assembly is one of the prevalent strategies used by living systems to fabricate ensembles of precision nanometer-scale structures and devices. The push for analogous approaches to create synthetic nanomaterials has led to the development of a large class of programmable crystalline structures. However, many applications require `self-limiting’ assemblies, which autonomously terminate growth at a well-defined size and geometry. For example, curved architectures such as tubules, vesicles, or capsids can be designed to self-close at a particular size, symmetry, and topology. But developing synthetic strategies for self-closing assembly has been challenging, in part because such structures are prone to polymorphism that arises from thermal fluctuations of their local curvature, a problem that worsens with increased target size. Here we demonstrate a strategy to eliminate this source of polymorphism in self-closing assembly of tubules by increasing the assembly complexity. In the limit of single-component assembly, we find that thermal fluctuations allow the system to assemble nearby, off-target structures with varying widths, helicities, and chirality. By increasing the number of distinct components, we reduce the density of off-target states, thereby increasing the selectivity of a user-specified target structure to nearly 100\%.  We further show that by reducing the design constraints by targeting either the pitch or the width of tubules, fewer components are needed to reach complete selectivity. Combining experiments with theory, our results reveal an economical limit, which determines the minimum number of components that are required to create arbitrary assembly sizes with full selectivity. In the future, this approach could be extended to more complex self-limited structures, such as shells or triply periodic surfaces.
\end{abstract}

\maketitle

\section*{Introduction}

The design and control of self-assembly pathways is a promising route for creating complex, functional nanomaterials~\cite{Whitesides2002Mar}. Recent successes in colloidal self-assembly focus primarily on synthesizing spatially unbounded, dense crystalline materials, or spatially-limited architectures, like clusters, membranes, and filaments, whose dimensions are specified by the building-block sizes~\cite{kraft2009colloidal,sacanna2013shaping,wang2014three,chen2011directed,he2020colloidal,Hensley2022self,Hensley2023macroscopic,zerrouki2008chiral,sacanna2010lock,chen2011supracolloidal,wolters2015self,tikhomirov2018triangular,oh2019colloidal,kahn2022encoding}. However, Nature is brimming with examples from a different class of structures that have self-regulated cavity sizes that are much larger than the size of the individual building blocks. Importantly, these self-limiting cavity sizes are essential to the functionality of various biological devices and materials, such as responsive containers that can selectively package specific genetic material, as in viral capsids~\cite{Caspar1962physical}, or photonic nanostructures, like those found in some butterfly wings that produce their structural coloration~\cite{michielsen2008gyroid}. Inspired by these examples, synthetic colloidal assembly has recently taken a large leap forward by developing new colloidal building blocks with directional binding, complex geometries, and specific interactions that can target assembly of similar self-limiting architectures, such as icosahedral shells and cylindrical tubules~\cite{Wagenbauer2017gigadalton,Sigl2021programmable,Hayakawa2022}.

\begin{figure*}[!th]
    \centering
    \includegraphics[width=\linewidth]{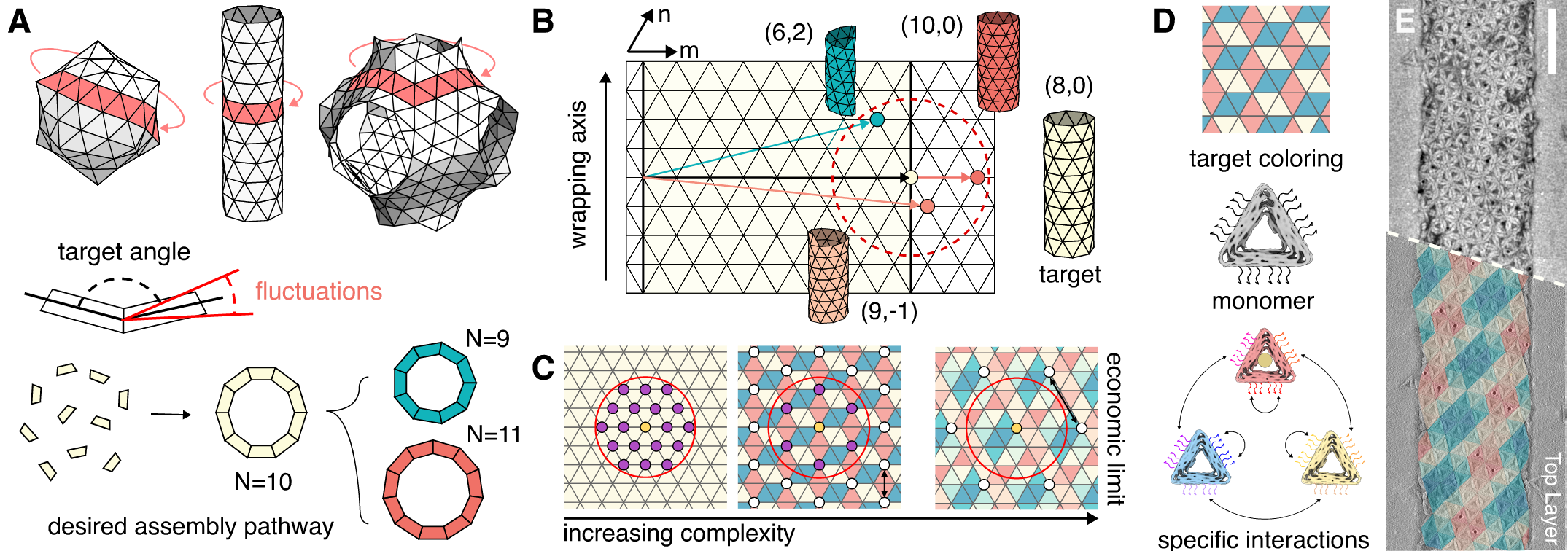}
    \caption{\textbf{Improving selectivity in self-closing assemblies}. (A) Examples of self-closing architectures---the unit cell of a P-surface (right), a cylinder (middle), and an icosahedral shell (left)---and closed loops of monomers (denoted in red). Fluctuations in the angles between subunits lead to a distribution of closure sizes in loops. (B) Schematic of closure fluctuations for a tubule. Tubules form from wrapping a sheet while satisfying a periodic boundary condition. Any discrete assembly can be represented by a unique vector between two vertices on the triangulated plane, corresponding to the circumference of that tubule. The pair of numbers ($m,n$) represents the numbers of steps in the $\mathbf{m}$ and $\mathbf{n}$ directions of the lattice to create this circumference vector. States accessible to thermal fluctuations around the target state lie within the dotted circle. Examples of target and off-target assemblies are shown. (C) Scheme for removing off-target states. Accessible states are shown as purple and states allowed by the coloring pattern are white. As the number of colors increases, the distance between similar vertices grows and eventually removes all allowed off-target states from the fluctuation area (red circle). (D) Experimental scheme for realizing colorings of tubules. Lines on the monomer edges denote ssDNA locations. Arrows between colored monomers show binding rules. (E) Example TEM micrograph for the 16-color tubule. The bottom shows a single layer of the tubule. 10-nm-diameter gold-nanoparticles identify the red component. Scale bar is 200 nm.}
    \label{Fig:fig1-scheme}
\end{figure*}

A common strategy used by Nature to assemble self-limiting structures exploits self-closure~\cite{Hagan2021equilibrium}, in which accumulated curvature between bound subunits allows the assembly to close upon itself during growth, thereby terminating assembly at a finite size. While translating this principle of self-closure could open new doors in synthetic colloidal assembly, self-closure has an associated fundamental challenge that must be solved to assemble precise, self-limiting architectures: Thermal fluctuations give rise to variations in the curvature of the growing assembly, leading to a distribution of the cavity sizes that result in the final structures~\cite{helfrich1986size}. This problem relates to the fact that the programmed angles between subunits lead to closed loops around the cavities, as seen for different discrete assemblies in Fig.~\ref{Fig:fig1-scheme}A. Though a certain average curvature may be targeted, the finite bending rigidity of the assemblies leads to a spread of sizes that these loops can form~\cite{Frey2020stochastic,Hagan2021equilibrium,Videbaek2022,Sui2010structural}. This aspect of assembly is easily seen in the context of forming a ring, Fig.~\ref{Fig:fig1-scheme}A, in which the collective fluctuations of many directional bonds can admit larger or smaller rings~\cite{Wagenbauer2017gigadalton}. The larger the number of subunits in the ring, the broader the distribution of states that can be accessed, and thus the ability to selectively assemble the target ring becomes more difficult with increasing ring size. Because the self-limiting cavity sizes are fundamental to how these types of structures function, developing generally applicable solutions to this challenge is essential to realizing synthetic self-limiting architectures with functionalities that rival their natural counterparts.

In this article, we revisit the initial promise of programmable assembly, and demonstrate a general strategy for eliminating polymorphism in self-closing assembly by increasing the assembly complexity. By combining theory and experiment we reveal an economical limit, in which the selectivity of a single target structure can be increased to 100\% using a minimal number of components. Our approach extends symmetry-based theories, such as that from Caspar and Klug~\cite{Caspar1962physical}, which identify the minimum number of components required to form a given structure. In particular, we demonstrate our strategy for the specific case of assembling cylindrical tubules from triangular DNA origami subunits (Fig.~\ref{Fig:fig1-scheme}B--E). We increase the complexity of our assemblies by incorporating a multi-component, periodic coloring pattern that specifies the arrangement of various components within the tubule. By using colorings with more types of components, fewer tubule geometries are commensurate with both the geometry of the tubule and the periodicity of the coloring, which reduces the number of off-target states that are thermally accessible. By borrowing ideas from two-dimensional addressable assembly~\cite{tikhomirov2018triangular,murugan2015undesired,Wintersinger2023multi}, we decouple the interactions between subunits from the geometry of the monomer to realize assemblies built from up to 16 unique components, demonstrating that our strategy allows assembly with essentially arbitrarily high complexity. Ultimately, we identify an economical threshold---which grows as the size of the cavity squared---where a single assembly state can be selected with a minimal amount of coloring complexity and rationalize this limit using a simple Helfrich energy model. We conclude by showing that one can target lower dimensional properties, like the width or the pitch, which reduces the constraints on assembly and thus provides a more economical scaling that is proportional to the self-limiting size.

\section*{Improving specificity using multiple components}

Our strategy works as follows. A cylindrical tubule can be conceptualized as a sheet that closes upon itself. Because you can tile a sheet with identical equilateral triangles, only one component is required~\cite{Hayakawa2022} and any tubule state can be identified with a unique pair of numbers ($m,n$) that corresponds to the shortest closed path around the tubule (directions $m$ and $n$ in Fig.~\ref{Fig:fig1-scheme}B). But having only a single component also allows the sheet to close on itself in many different ways since all vertices of the triangulation are identical, Fig. 1B. While a single tubule state can be preferred, as specified by the assembly's curvature, the finite bending rigidity of the sheet could admit many neighboring tubules, with similar curvatures and therefore similar bending energies. These accessible geometries can be thought of as off-target states that occupy an area of vertices around some ground-state vertex. The challenge of eliminating polymorphism in the final structures then boils down to removing all undesired states from within this area. 

To accomplish this goal, we color triangles in a periodic way using an increasing number of colors, which removes translational symmetries of the sheet and reduces the number of allowed closure states, Fig. 1C. As additional complexity of coloring is used, the allowed closure states are pushed further apart and eventually there are no allowed vertices within the area of fluctuations. At this point, only a single accessible assembly state is permitted by the matching rules.

\begin{figure*}[!t]
    \centering
    \includegraphics[width=\linewidth]{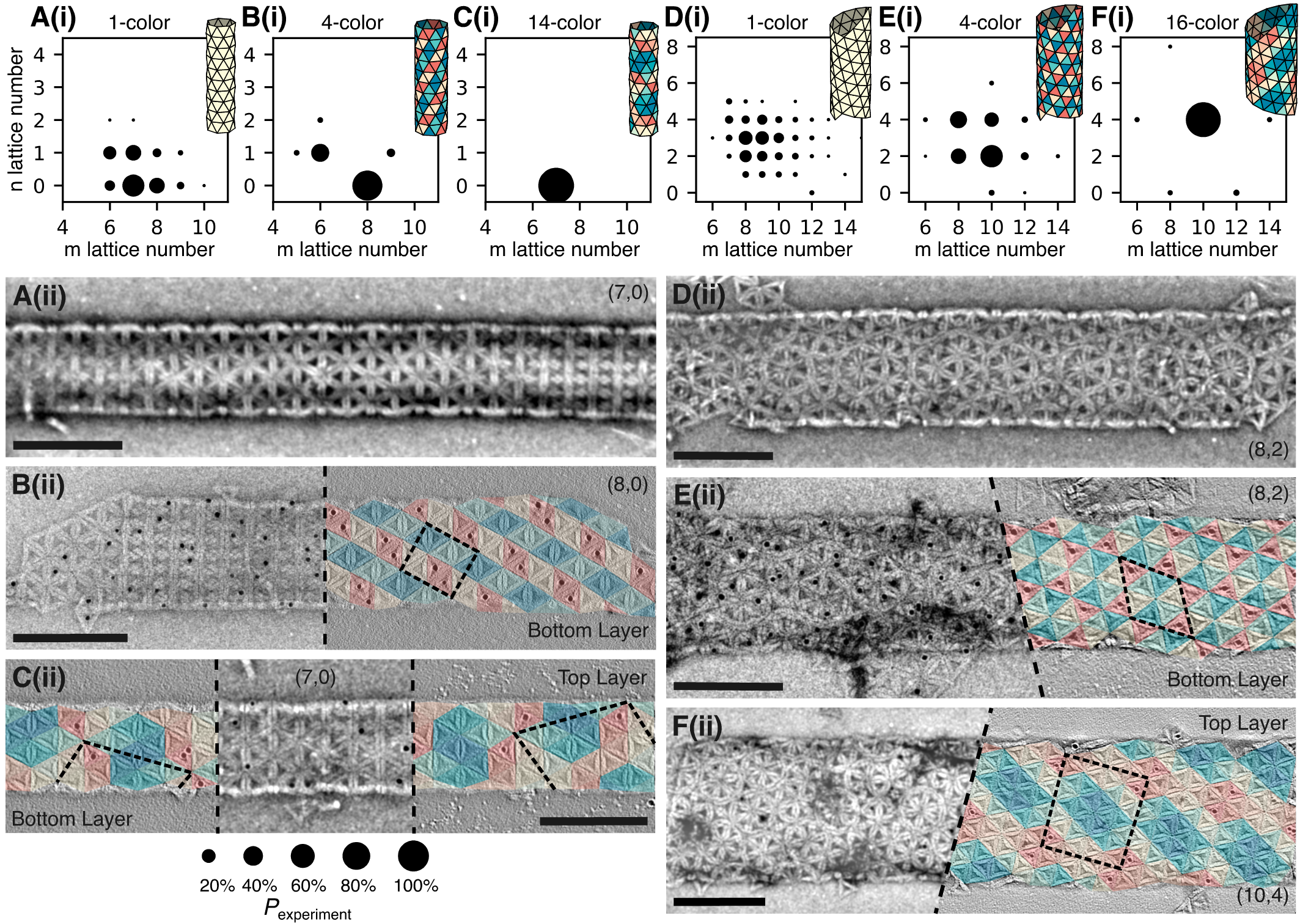}
    \caption{ \textbf{Increased selectivity with increased complexity}. (A,B,C) (i) Probability distributions of different tubule types for 1-, 4-, and 14-color assemblies with the (6,0) monomer. Each circle denotes a single ($m, n$) tubule type and the area of each point denotes the relative probability. Illustrations of the most probable tubule for each distribution are shown with each plot. (ii) Representative TEM images of (6,0) tubule assemblies with the 1-, 4-, and 14-color designs. (D,E,F) (i) Probability distributions for 1-, 4-, and 16-color assemblies with the (10,0) monomer. (ii) Representative TEM images of (10,0) tubule assemblies with the 1-, 4-, and 16-color designs. Tubules in B(ii),C(ii),E(ii), and F(ii) have the corresponding tiling overlayed with a single color labeled with 10-nm-diameter gold nanoparticles. Only positive $n$ points are shown since single TEM images cannot differentiate between left and right-handed structures. All scale bars are 200 nm.}
    \label{Fig:fig2-specificity}
\end{figure*}

We develop an experimental system to test this concept, comprised of triangular subunits made via DNA origami, that targets a user-specified cylindrical tubule by controlling the interaction specificity and the dihedral angles between neighboring subunits~\cite{Hayakawa2022}. To encode the dihedral angles, we bevel each side of the triangle. The monomers also need specific interactions to preserve their orientation with respect to the tubule axis. To realize specific interactions, we place six single-stranded DNA (ssDNA) segments with six-base-long binding domains along each edge (Fig.~\ref{Fig:fig1-scheme}D). Using sticky-end hybridization allows us to encode many low cross-talk interactions~\cite{Seeman1990,Wu2012Polygamous,Tikhomirov2017fractal,Wintersinger2023multi, kahn2022encoding} without changing any internal routing of the DNA origami, which could otherwise have unintended effects on the monomer structure~\cite{Wagenbauer2017, Sigl2021programmable,Hayakawa2022}. Finally, we assemble tubules at a constant temperature chosen such that the intersubunit interactions are weak and reversible.

We find that a single component assembles a distribution of tubules with varying width and helcity. We first design a triangular monomer that targets a (6,0) tubule and classify the tubules that assemble with transmission electron microscopy (TEM). In particular, we measure both the width and pitch of each tubule to identify its ($m,n$) type, and then construct a distribution of states, Fig.~\ref{Fig:fig2-specificity}A. The distribution reveals that the monomer prefers to form (7,0) tubules---close to our target of (6,0)---but that over half of the observed tubules form other assembly states distributed near the preferred state, including both achiral and chiral tubules. 

To circumvent the formation of off-target states, we engineer the free-energy landscape around the preferred state by building the tubule from a periodic arrangement of a larger number of components. 
More specifically, we generate a matrix of pairwise subunit interactions by finding periodic colorings of the plane for varying numbers of colors, Fig.~\ref{Fig:fig1-scheme}C (Suppl. Sec. II). The set of adjacent colors specifies which unique, specific interactions are required to assemble that pattern, which we implement by designing the sticky-end sequences of our origami subunits. Each color then corresponds to a subunit with a unique set of intersubunit interactions. Going forward, we refer to a component as a `color'. 

By choosing colorings that permit the preferred (7,0) state, we aim to reduce the density of nearby, undesired assemblies. Figures~\ref{Fig:fig2-specificity}B and C show the distributions of 4- and 14-color assemblies. As the number of colors, $N_\mathrm{colors}$, increases we find that the density of available states decreases and the probability of the most likely tubule state, which we call `selectivity', increases. Importantly, images of individual tubules show that the quality of assemblies does not diminish as $N_\mathrm{colors}$ increases, Fig.~\ref{Fig:fig2-specificity}B(ii) and C(ii). We also confirm the specificity of the intersubunit interactions by labeling a single color with 10-nm-diameter gold nanoparticles and performing TEM tomography. While we cannot unambiguously infer the colors of all of the components, we can conclude that the pattern of gold nanoparticles is consistent with our designed coloring in all cases.  

\begin{figure}[!t]
    \centering
    \includegraphics[width=\linewidth]{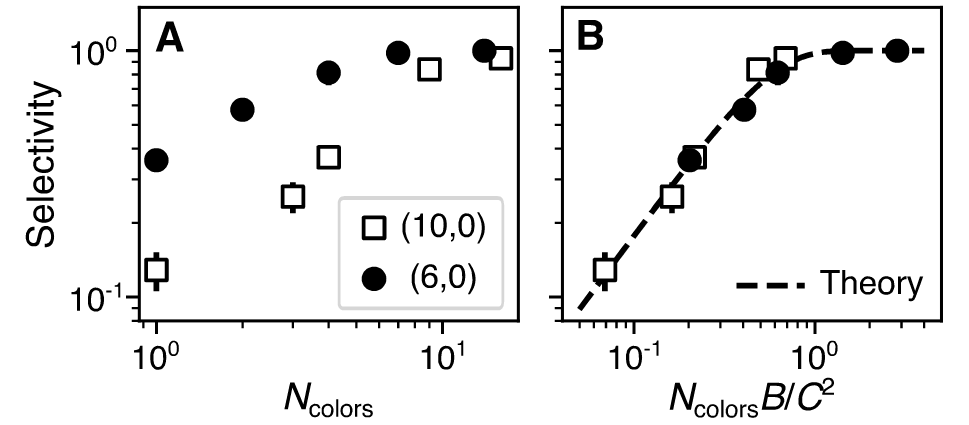}
    \caption{ \textbf{Scaling for increased selectivity}. (A) Probability of the most common assembly state, called selectivity, against $N_\mathrm{colors}$. (B) Selectivity rescaled by the fluctuation area $N_\mathrm{colors}B/C^2$. $C$ values are taken from the most probable tubule state. Fitting the single-component assemblies to the Helfrich model provides estimated values of the bending rigidity of 10.0 \kT\  and 6.7 \kT\ for the (6,0) and (10,0) monomers, respectively. See Suppl. Sec. VI and Reference~\cite{Videbaek2022} for a description of the theory.}
    \label{Fig:fig3-SpecificTheory}
\end{figure}

Because fluctuations of the dihedral angles between adjacent subunits give rise to the breadth of states, we expect that targeting a larger diameter tubule will result in a broader distribution of off-target structures and thus will require a greater degree of complexity to achieve full selectivity. To test this hypothesis we design a second DNA origami monomer with bevel angles targeting a (10,0) tubule. Figure~\ref{Fig:fig2-specificity}D shows the distribution of states for this monomer and an example of an assembled tubule. As anticipated, the number of off-target states increases in comparison to the (6,0) tubule. Again, we can improve selectivity through increased assembly complexity. Figures~\ref{Fig:fig2-specificity}E and F show 4- and 16-color assemblies and a corresponding increase in the assembly selectivity. As before, we see that the quality of tubules is preserved with increasing complexity and that the interaction specificity is consistent with our designed colorings, as seen in the TEM images.

We use a simple Helfrich model to relate the distribution of available states to the physical properties of the assembly~\cite{Helfrich1988,Videbaek2022,Fang2022}. The model assumes that tubule assembly begins by the formation of a circular patch, and thus the area of states that can form around the target due to thermal fluctuations is $C^2/B$, where $C$ is the preferred circumference non-dimensionalized by the monomer edge length and $B$ is the bending rigidity non-dimensionalized by \kT\ (Suppl. Sec. VI). This area is independent of the assembly complexity. Our coloring patterns introduce another lengthscale: the distance between similar vertices. This length grows with the size of the primitive cell of the coloring, whose area increases as $N_\mathrm{colors}$ (Fig.~\ref{Fig:fig1-scheme}C). When these two areas are comparable, we expect that the nearest allowed off-target assembly is outside the area accessible by thermal fluctuations. Thus, we expect a crossover to 100\% selectivity when 
\begin{equation}
    N_\mathrm{colors}B/C^2 \gtrsim 1.
\end{equation}

Our data of the selectivity collapse when replotted according to our Helfrich model. In particular, Fig.~\ref{Fig:fig3-SpecificTheory} shows that the selectivity increases linearly with an increasing number of colors and plateaus to full selectivity when $N_\mathrm{colors}B/C^2 \approx 1$. This crossover point corresponds to the most economical way to target a single assembly state, after which additional complexity provides no further benefit.

\section*{Limiting lower-dimensional properties of assemblies}

\begin{figure}[!tb]
    \centering
    \includegraphics[width=\linewidth]{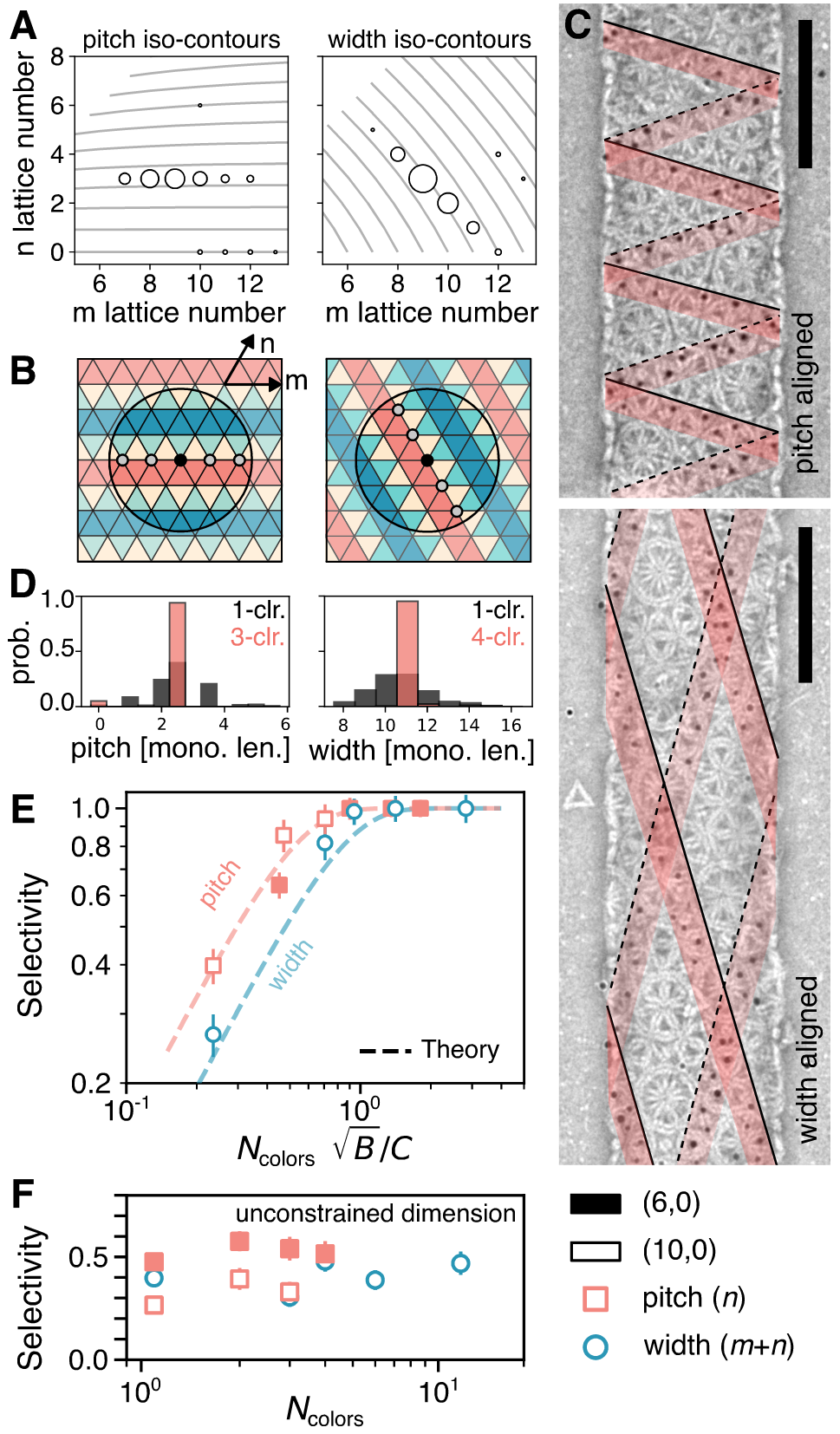}
    \caption{  \textbf{Limiting single degrees of freedom in assembly}. (A) Iso-contours of the pitch and width for tubules for varying $m$ and $n$. (B) Schematic of allowed states for linear colorings for pitch or width alignment and (C) TEM images of 4-color width- and pitch-controlled tubules with the (10,0) monomer. Lines show seams of labeled particles (solid and dashed lines denote the top and bottom layers of the tubule). (D) Probability distributions for the pitch and width of multi-component tubules compared to one-color tubules. Units for the pitch and width are the edge length of a monomer. Complete ($m,n$) distributions of these assemblies are shown in (A). (E) Selectivity of tubules that have the same $m+n$ (width-control, circles) or the same $n$ (pitch-control, squares). The dashed lines are theoretical predictions from Fig. S5. (F) Selectivity of the unconstrained dimension versus $N_\mathrm{colors}$. }
    \label{Fig:fig4-width-pitch}
\end{figure}

While the promise of complete selectively is tantalizing, the scaling of the number of components with the square of the self-limited length becomes daunting for larger structures. Therefore, we explore the possibility that targeting a lower-dimensional quantity, like the width or pitch, could be accomplished using fewer components. For each tubule geometry, i.e., each ($m,n$), there is a unique width and pitch of the assembly. But there also exist families of tubules that have nearly constant width or pitch. Figure~\ref{Fig:fig4-width-pitch}A shows iso-contours of these properties in ($m,n$) space. We note that lines of constant $n$ or $m+n$ approximate lines of constant pitch or width, respectively. Therefore, we are able to target families of nearly constant width or pitch using linear colorings, in which seams of the same color lie along a single lattice direction, by aligning them along either $\mathbf{n}$ or $\mathbf{m-n}$ directions, Fig.~\ref{Fig:fig4-width-pitch}B.

We validate this scheme in experiment and confirm that it yields tubule distributions that are highly selective in their width or pitch. By labeling a single color with gold nanoparticles we can see that lines of the same color correspond to lattice directions that are closest to the circumferential or axial directions for the pitch- and width-controlled tubules, respectively (Fig.~\ref{Fig:fig4-width-pitch}C). We also compare the distributions of tubule widths and pitch for two of our (10,0) tubule experiments against a single-color experiment, and find that our linear colorings create assemblies with tightly peaked distributions for the desired property, Fig.~\ref{Fig:fig4-width-pitch}D (Suppl. Fig.~S9).

Using a similar argument from the Helfrich model, we find that placing reduced constraints on the assembly leads to an economical threshold that scales linearly with the self-limited length scale, rather than quadratically as before. Since we target seams of similar states, we are now concerned not with the area of thermal fluctuations, but with their linear extent, $C/\sqrt{B}$. For linear colorings, the separation between the same component type grows as $N_\mathrm{colors}$. 

To compare our experimental data to the Helfrich model, we plot the selectivity of tubule states with the same $n$ or $m+n$ against $N_\mathrm{colors}\sqrt{B}/C$. Figure~\ref{Fig:fig4-width-pitch}E shows that our data collapse in agreement with the Helfrich model, with the economical point occurring when $N_\mathrm{colors}\sqrt{B}/C\approx1$. However, we find that reducing the design constraints comes at the cost that the unconstrained dimension (e.g. the pitch when one targets the width) does not see any improvement in selectivity for increasing complexity, Fig.~\ref{Fig:fig4-width-pitch}F.

\begin{figure}[!t]
    \centering
    \includegraphics[width=\linewidth]{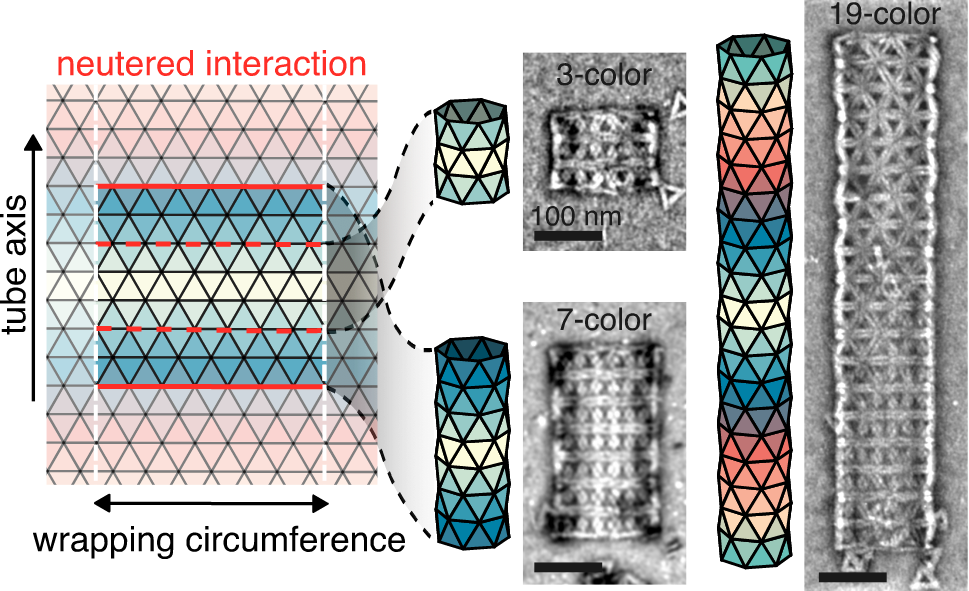}
    \caption{  \textbf{Tubule length limitation}. For achiral tubules (such as those made by the (6,0) monomer), a linear coloring constrains tubules to length $n$ by neutering one edge of the $n$-th color. Dashed and solid red lines denote the neutered interactions for $n=$3 and 7. TEM images for 3-, 7-, and 19-color assemblies are shown. All scale bars are 100 nm.}
    \label{Fig:fig5-tubelets}
\end{figure}

Whereas we have so far focused on constraining the ($m,n$) states that can form, we conclude by showing how assembly complexity can also be used to constrain a different aspect of the tubule geometry: their length. We use the same class of linear colorings as above, but now we passivate a single edge of one color to prevent the further addition of monomers along that seam, Fig.~\ref{Fig:fig5-tubelets}. In this way, the number of colors constrains the tubule length. In Fig.~\ref{Fig:fig5-tubelets} we show representative TEM micrographs of our (6,0) monomer with 3-, 7-, and 19-color assemblies. In all cases, we find that the system reaches an equilibrium state characterized by a mixture of monomers, small clusters, and finite, $n$-length assemblies (Suppl. Sec. VII). The spirit of this type of constraint is different than those mentioned previously since we are not increasing the complexity to reduce the number of states accessible by thermal fluctuations, but rather to cap the growth of assemblies, highlighting the various ways that assembly complexity can be used to engineer self-limitation.

\section*{Conclusions}

In this work, we have demonstrated a scheme for preventing off-target assemblies by increasing the complexity of the initial assembly mixture. By imposing complex colorings, implemented through matrices of specific intersubunit attractions, assembly states that were accessible due to fluctuations in the curvature become disallowed, leading to full selectivity of the target geometry. By imposing different numbers of constraints on the assembly, we were able to either achieve full selectivity for both the pitch and width of tubules (Fig.~\ref{Fig:fig2-specificity} and ~\ref{Fig:fig3-SpecificTheory}) or either one separately (Fig.~\ref{Fig:fig4-width-pitch}). The benefit of selecting a single dimension of the structure, say the width, is that it can be done using significantly fewer colors, which becomes essential as the self-limiting dimension becomes arbitrarily large relative to the subunit size. Unfortunately, targeting the pitch or width alone comes with a lack of selectivity for the unconstrained dimension, e.g. the pitch. We show examples of this trade-off in Suppl. Fig.~S8. While in the limit of large target sizes selectivity eventually falls off, we showed that increasing the complexity mitigates the marginal losses of selectivity, enhanced by a factor of $N_\mathrm{colors}$ over the standard Helfrich expectation. 

We anticipate that our strategy is generalizable to other self-closing architectures. A useful point of comparison is the Caspar-Klug (CK) construction for icosahedral capsids~\cite{Caspar1962physical}. By utilizing the symmetry of icosahedral shells, the CK rules provide a method for determining the minimal number of components needed to form a shell of arbitrary size. However, for too low of a bending modulus, capsids can assemble asymmetric, defective structures. These defects often come from disclinations that form at 5-fold or 3-fold points. By increasing the number of colors around certain symmetry points in the structure, beyond the minimum prescribed by CK theory, we anticipate that strategies like the one we present here could reduce the propensity for these types of systems to form defects.

Despite the great improvement in specificity that we see using multi-color assemblies, there are still challenges that need to be overcome. The foremost is the kinetics of growth. The timescale for assembly increases with increasing complexity since the chance of the correct color monomer finding a binding site at the growth front decreases with $N_\mathrm{colors}$ (Suppl. Sec. VIII). This decrease in assembly rate could be compensated for by increasing the monomer concentration, but this strategy could become costly. Instead, future strategies may exploit hierarchical pathways to assembly both for accelerating assembly and possibly reducing the desired complexity if certain kinetic paths can be found that disallow off-target structures. An important aspect of the kinetics that this work reveals is that assembly near the economical point is favorable. In the 14-color assembly with the (6,0) monomer, even though all closed structures formed the same ($m,n$) state, a small fraction of assemblies missed the point of closure and overgrew as a sheet that wrapped around itself like a scroll. We hypothesize that this type of misassembly will become even more prevalent if the assembly complexity passes the point of optimal economy, $N_\mathrm{colors}B/C^2\gg 1$. 

An important consideration with all types of multi-component assemblies is the possibility of crosstalk between different components that can stabilize unintended contacts~\cite{murugan2015undesired, huntley2016information}. In this context, we can ask why our designs worked so well and do not seem to suffer from the inevitability of cross-talk between a limited set of DNA sequences~\cite{Wu2012Polygamous}? Here we highlight that the site addressability of DNA origami allows for the inclusion of interaction geometry as an additional knob for tuning the specificity of interactions~\cite{Wintersinger2023multi}. In fact, for some of our interaction patterns, we reused strands from other interactions but altered their arrangement on the faces of the interacting edges. This combination of sequence specificity and spatial commensurability greatly increases the number of unique interactions that can be specified by a finite library of DNA sequences. Furthermore, increasing the number of sticky ends per side could be used to increase this combinatoric aspect further, allowing for many more components than have been explored here.

Coloring, the process of finding allowed, distinguishable identities for particles in a structure, is an attractive way to increase the complexity of assemblies, though the inverse design challenge of finding an appropriate coloring for a desired outcome is not always straightforward. Here we were able to exploit the symmetries and periodicity of the tubule to generate any desired coloring. More complex geometries, such as capsids or gyroids, require considering local point symmetries of the structure~\cite{Caspar1962physical,Sigl2021programmable,tanaka2023programmable,duque2023limits} to encode the correct interactions, while fully addressable clusters often reveal their allowed colorings through detailed searches of all possible interacitons~\cite{Zeravcic2014size}. Recent efforts have also shown that colorings can be used to encode hierarchical structures into crystalline self-assemblies~\cite{kahn2022encoding}. As assembly structures become more complex, finding the right scheme to encode large libraries of interactions may require new methods, such as SAT-assembly~\cite{russo2022sat, Sciortino2023design}. Additionally, it will be equally important to be able to predict and thus program the right binding free energies to facilitate robust assembly~\cite{jacobs2015rational,curatolo2022assembly, bupathy2022temperature}.

Going forward, our scheme could be expanded to systems with subunits with both unique interactions and unique geometries to target a wider range of self-limiting architectures. Whereas we focused on a base monomer with fixed geometry, other target structures have the need for varying subunit geometries, as with icosahedral shells~\cite{Sigl2021programmable} or for surfaces with varying Gaussian curvature, such as helical cylinders, toroids, or open crystalline structures~\cite{duque2023limits}. Moreover, many studies have recently demonstrated that DNA origami can be dynamically reconfigured~\cite{liu2020modular, zhang2019dynamic, song2017reconfiguration, kim2023harnessing}, so it is within sight to imagine a base monomer with adjustable edge lengths or bevel angles that could be used to produce new nanoscale devices. \\

\begin{acknowledgments}
We thank Berith Isaac and Amanda Tiano for their technical support with electron microscopy. TEM images were prepared and imaged at the Brandeis Electron Microscopy facility. This work is supported by the Brandeis University Materials Research Science and Engineering Center (MRSEC) (NSF DMR-2011846) and the National Science Foundation (NSF DMR-2309635). D.H. acknowledges support from the Masason Foundation. W.B.R. acknowledges support from the Smith Family Foundation.
\end{acknowledgments}

\bibliography{main.bib}

\end{document}


\title{Supplemental Information for ``Economical routes to size-specific assembly of self-closing structures''}

\author{Thomas E. Videb{\ae}k}
\affiliation{Martin A Fisher School of Physics, Brandeis University, Waltham, MA, 02453, USA}
\author{Daichi Hayakawa}
\affiliation{Martin A Fisher School of Physics, Brandeis University, Waltham, MA, 02453, USA}
\author{Gregory M. Grason}
\affiliation{Department of Polymer Science and Engineering, University of Massachusetts, Amherst, Massachusetts 01003, USA}
\author{Michael F. Hagan}
\affiliation{Martin A Fisher School of Physics, Brandeis University, Waltham, MA, 02453, USA}
\author{Seth Fraden}
\affiliation{Martin A Fisher School of Physics, Brandeis University, Waltham, MA, 02453, USA}
\author{W. Benjamin Rogers}
\affiliation{Martin A Fisher School of Physics, Brandeis University, Waltham, MA, 02453, USA}

\maketitle

\setcounter{figure}{0}
\makeatletter 
\renewcommand{\thefigure}{S\arabic{figure}}

\section{Materials and Methods}

\textbf{Folding DNA origami.} To assemble our DNA origami monomers, we make a solution with 50 nM of p8064 scaffold (Tilibit), 200 nM of each staple strand (Integrated DNA Technologies [IDT]; Nanobase structures 234 and 235~\cite{Nanobase} for sequences), and 1x folding buffer. We then anneal this solution using a temperature protocol described below. Our folding buffer, from here on referred to as FoBX, contains 5~mM  Tris Base, 1~mM EDTA, 5~mM NaCl, and X~mM MgCl$_2$. We use a Tetrad (Bio-Rad) thermocycler to anneal our samples.

To find the best folding conditions for each sample, we follow a standard screening procedure to search multiple MgCl$_2$ concentrations and temperature ranges~\cite{Wagenbauer2017,Hayakawa2022}, and select the protocol that optimizes the yield of monomers while limiting the number of aggregates that form. All particles used in this study were folded at 17.5~mM MgCl$_2$ with the following annealing protocol: (i) hold the sample at 65~$^\circ$C for 15 minutes,  (ii) ramp the temperature from 58~$^\circ$C to 50~$^\circ$C with steps of 1~$^\circ$C per hour, (iii) hold at 50~$^\circ$C until the sample can be removed for further processing. \\

\textbf{Agarose gel electrophoresis.} We use agarose gel electrophoresis to assess the folding protocols and purify our samples with gel extraction. We prepare all gels by bringing a solution of 1.5\%~(w/w) agarose in 0.5X TBE to a boil in a microwave. Once the solution is homogenous, we cool it to 60~$^\circ$C using a water bath. We then add MgCl$_2$ and SYBR-safe (Invitrogen) to have concentrations of 5.5~mM MgCl$_2$ and 0.5x SYBR-safe. We pour the solution into an Owl B2 gel cast and add gel combs (20 $\upmu$L wells for screening folding conditions or 200~$\upmu$L wells for gel extraction), which cools to room temperature. A buffer solution of 0.5x TBE and 5.5~mM MgCl$_2$, chilled at 4~$^\circ$C for an hour, is poured into the gel box. Agarose gel electrophoresis is run at 110 V for 1.5--2 hours in a 4~$^\circ$C cold room. We scan the gel with a Typhoon FLA 9500 laser scanner (GE Healthcare) at 100~$\upmu$m resolution. \\

\textbf{Sample purification.} After folding, we purify our DNA origami particles to remove all excess staples and misfolded aggregates using gel purification. If the particles have self-complementary interactions, they are diluted 2:1 with 1xFoB2 and held at 47~$^\circ$C for 30 minutes to unbind higher-order assemblies. The folded particles are run through an agarose gel (now at a 1xSYBR-safe concentration for visualization) using a custom gel comb, which can hold around 2~mL of solution per gel. We use a blue fluorescent light table to identify the gel band containing the monomers. The monomer band is then extracted using a razor blade. We place the gel slices into a Freeze ’N Squeeze spin column (Bio-Rad), freeze it in a -20~$^\circ$C freezer for 5 minutes, and then spin the solution down for 5~minutes at 12~krcf. The concentration of the DNA origami particles in the subnatant is measured using a Nanodrop (Thermo Scientific). We assume that the solution consists only of monomers, where each monomer has 8064 base pairs.

Since the concentration of particles obtained after gel purification is typically not high enough for assembly, we concentrate the solution using ultrafiltration~\cite{Wagenbauer2017}. First, a 0.5-mL Amicon 100-kDa ultrafiltration spin column (Millipore) is equilibrated by centrifuging down 0.5~mL of 1xFoB5 buffer at 5~krcf for 7~minutes. Then, the DNA origami solution is added and centrifuged at 14 krcf for 15 minutes. We remove the flow-through and repeat the process until all of the DNA origami solution is filtered. Finally, we flip the filter upside down into a new Amicon tube and spin down the solution at 1~krcf for 2~minutes. The concentration of the final DNA origami solution is then measured using a Nanodrop. \\

\textbf{Tubule assembly.} Assembly experiments are conducted with DNA origami particle concentrations ranging from 2~nM to 30~nM. For assemblies that are made up of multiple colors, the quoted concentration is the total concentration of all subunits, e.g. for a 10-nM experiment with $N$ colors, each color has a concentration of 10/$N$ nM. Assembly solutions have volumes up to 50~$\upmu$L with the desired DNA origami concentration in a 1xFoB20 buffer. The solution is placed in a 200~$\upmu$L PCR tube and loaded into a thermocycler (Bio-Rad), which is held at a constant temperature, ranging between 30 $^\circ$C and 50~$^\circ$C. The thermocycler lid is held at 100~$^\circ$C to prevent condensation of water on the cap of the PCR tube.

Since DNA hybridization is highly sensitive to temperature, we expect that there should be a narrow range of temperatures over which the system can assemble by monomer addition. To make sure that we assemble tubules within this regime, we prepare many samples over a broad range of temperatures. At high temperatures, we find that there are no large assemblies, implying that we are above the melting transition for our ssDNA interactions. As we lower the temperature, we find a transition to the formation of assembled tubules, but with increasing defect density as the temperature decreases. For this reason, all assembly experiments are conducted just below the melting transition. \\

\textbf{Labeling tubules with gold nanoparticles.} We first attach thiol-modified ssDNA (5'-HS-C$_6$H$_{12}$-TTTTTAACCATTCTCTTCCT-3', IDT) to 10-nm-diameter gold nanoparticles (AuNP) (Ted Pella) using a protocol similar to that in ref.~\cite{Sun2020valence}. We first reduce the thiolated strands using tris(2-carboxyethyl) phosphine (TCEP) solution (Sigma-Aldrich) by holding a mixture of 10~mM TCEP (pH 8) and 100~$\upmu$M thiol-DNA at room temperature for one hour on a vortex shaker. We remove excess TCEP with a 10-kDa Amicon filter in three washes of a 50~mM HEPES buffer (pH 7.4); we follow this with filter centrifugation at 4~krcf for 50~min at 4~$^\circ$C. After purification, we store thiolated DNA strands at -20~$^\circ$C until needed. To attach thiolated DNA to AuNPs, we mix DNA with AuNPs at a ratio of 300:1 in a 1x borate buffer (Thermo Scientific) and rotate the mixture at room temperature for 2~hours. After incubation, we increase the salt concentration in a stepwise manner to 0.25~M NaCl using a 2.5~M NaCl solution in five steps. After each salt addition, we rotate the AuNP solution at room temperature for 30~min. After the last addition, we let the AuNP solution age in the rotator overnight. To remove excess thiol-DNA strands, we wash the DNA-AuNP conjugates four times by centrifugation using a 1x borate buffer with 0.1~M NaCl. In each wash step, we centrifuge the DNA-AuNP solutions at 6.6~krcf for 1 hour. After the last wash, we measure the DNA-AuNP concentration using a Nanodrop, and store the solution at 4~$^\circ$C.

To attach AuNPs to tubules, we incorporate handles on the interior edges of the DNA origami subunit with a complementary sequence (5'-AGGAAGAGAATGGTT-3', IDT) to the DNA on the AuNP. For a multi-component assembly, only one subunit type has handles that bind to the AuNPs. After tubules have been assembled, we dilute the assembly solution into a mixture with final concentrations of 1~nM DNA origami monomers and 2~nM AuNP in 1xFoB20 and incubate at 32~$^\circ$C overnight. After incubation, samples are ready to be prepared for imaging. \\

\textbf{Fluorescence microscopy.} We incubate our DNA origami tubules with YOYO-1 dye (Invitrogen) at room temperature for a minimum of half an hour in a solution of 5 nM monomers, 500 nM YOYO-1, and 1xFoB at the MgCl$_2$ concentration of the assembly. This ratio of YOYO-1 to DNA origami is chosen so that there are 100 dye molecules per structure, a limit in which the dye’s impact on the structural integrity of the origami should be negligible~\cite{Gunther2010mechanical}. 1.6 $\upmu$L of the solution is pipetted onto a microscope slide that has been cleaned with Alconox, ethanol (90\%), acetone, deionized water, and subsequently plasma-cleaned. After deposition, a plasma-cleaned coverslip is placed on the droplet at an angle and carefully lowered so that the liquid film is as thin as possible. We find that this reduces the sample thickness to about the width of a tubule without damaging the tubules, allowing them to lie flat on the surface. Samples are imaged on a TE2000 Nikon inverted microscope with a Blackfly USB3 (FLIR) camera. \\

\textbf{Negative-stain TEM.} We first prepare a solution of uranyl formate (UFo). We boil doubly distilled water to deoxygenate it and then mix in UFo powder to create a 2\%~(w/w) UFo solution. We cover the solution with aluminum foil to avoid light exposure and vortex it vigorously for 20~minutes, after which we filter the solution with a 0.2~$\upmu$m filter. Lastly, we divide the solution into 0.2~mL aliquots, which are stored in a -80 $^\circ$C freezer until further use.

Before each negative-stain TEM experiment, we take a 0.2~mL UFo aliquot out from the freezer to thaw at room temperature. We add 4~$\upmu$L of 1~M NaOH and vortex the solution vigorously for 15~seconds. The solution is centrifuged at 4~$^\circ$C and 16~krcf for 8~minutes. We extract 170~$\upmu$L of the supernatant for staining and discard the rest.

The EM samples are prepared using FCF400-Cu grids (Electron Microscopy Sciences). We glow discharge the grid prior to use at -20~mA for 30 seconds at 0.1 mbar, using a Quorum Emitech K100X glow discharger. We place 4 $\upmu$L of the sample on the carbon side of the grid for 1 minute to allow adsorption of the sample to the grid. During this time, 5 $\upmu$L and 18 $\upmu$L droplets of UFo solution are placed on a piece of parafilm. After the adsorption period, the remaining sample solution is blotted on 11 $\upmu$m Whatman filter paper. We then touch the carbon side of the grid to the 5 $\upmu$L drop and blot it away immediately to wash away any buffer solution from the grid. This step is followed by picking up the 18 $\upmu$L UFo drop onto the carbon side of the grid and letting it rest for 30 seconds to deposit the stain. The UFo solution is then blotted and any excess fluid is vacuumed away. Grids are allowed to dry for a minimum of 15 minutes before insertion into the TEM. 

We image the grids using an FEI Morgagni TEM operated at 80 kV with a Nanosprint5 CMOS camera (AMT). The microscope is operated at 80 kV and images are acquired between x8,000 to x20,000 magnification. \\

\textbf{TEM tomography.} To obtain a tilt-series, we use an FEI F20 equipped with a Gatan Ultrascan 4kx4k CCD camera, operated at 200 kV. The grid is observed at x18000 magnification from -50~degrees to 50~degrees in 2-degree increments. The data is analyzed and the $z$-stack is reconstructed using IMOD~\cite{kremer_computer_1996}.\\

\textbf{Cryo-electron microscopy.} Higher concentrations of DNA origami are used for cryo-EM grids than for assembly experiments. To ensure that particles remain isolated from each other in the ice, we use passivated monomers, which have no ssDNA strands protruding from the faces of the DNA origami. To prepare samples, we fold between 1--2 mL of the folding mixture (50 nM scaffold concentration), gel purify it, and concentrate the sample by ultrafiltration, as described above. EM samples are prepared on glow-discharged C-flat 1.2/1.3 400 mesh grids (Protochip). Plunge-freezing of grids in liquid ethane is performed with an FEI Vitrobot with sample volumes of 3~$\upmu$L, blot times of 16~s, a blot force of -1, and a drain time of 0~s at 20~$^\circ$C and 100\% humidity.

Cryo-EM images for the (10,0) DNA origami monomer are acquired with the FEI Tundra TEM with a field emission gun electron source operated at 100 kV and equipped with an FEI Falcon II direct electron detector at a magnification of x59000. Single-particle acquisition is performed with SerialEM. The defocus is varied from -0.5 $\upmu$m to -4 $\upmu$m with a pixel size of 2.023 Angstrom.

Cryo-EM images for the (6,0) DNA origami monomer are acquired with a Tecnai F30 TEM with a field emission gun electron source operated at 300 kV and equipped with an FEI Falcon II direct electron detector at a magnification of x39000. Single-particle acquisition is performed with SerialEM. The defocus is set to -2 $\upmu$m for all acquisitions with a pixel size of 2.87 Angstrom. \\

\textbf{Single-particle reconstruction.} \noindent 
Image processing is performed using RELION-3~\cite{Zivanov2018RELION3}. Contrast-transfer-function (CTF) estimation is performed using CTFFIND4.1~\cite{Rohou2015ctffind}. After picking single particles, we perform a reference-free 2D classification from which the best 2D class averages are selected for processing, estimated by visual inspection. The particles in these 2D class averages are used to calculate an initial 3D model. A single round of 3D classification is used to remove heterogeneous monomers and the remaining particles are used for 3D auto-refinement and post-processing. Figures~\ref{Sfig:monomerCryo-6-0} and \ref{Sfig:monomerCryo-10-0} show views of the reconstructions and the resolution curves. The post-processed maps are deposited in the Electron Microscopy Data Bank with entry EMD-43226 and EMD-43227. \\

\section{Generating coloring patterns for tubules} \label{SIsec:tilingPatterns}

To know what interactions to program between our monomer species, we first need to generate coloring patterns. A coloring pattern is the assignment of color to each triangle on the plane such that it can be wrapped into a tubule. This restriction means that the coloring needs to have translational symmetry and can have at most 2-fold symmetry since tubules have a well-defined axis (these criteria correspond to either `o' or `2222' symmetry groups on the plane). To minimize the number of colors needed to create a tubule, we only use patterns with 2-fold symmetry in this work. Our previous study~\cite{Videbaek2022} used a brute-force computational search over all possible interactions between subunits to generate allowed patterns, but this strategy becomes intractable for more than ten colors.

\begin{figure*}[!b]
    \centering
    \includegraphics[width=\linewidth]{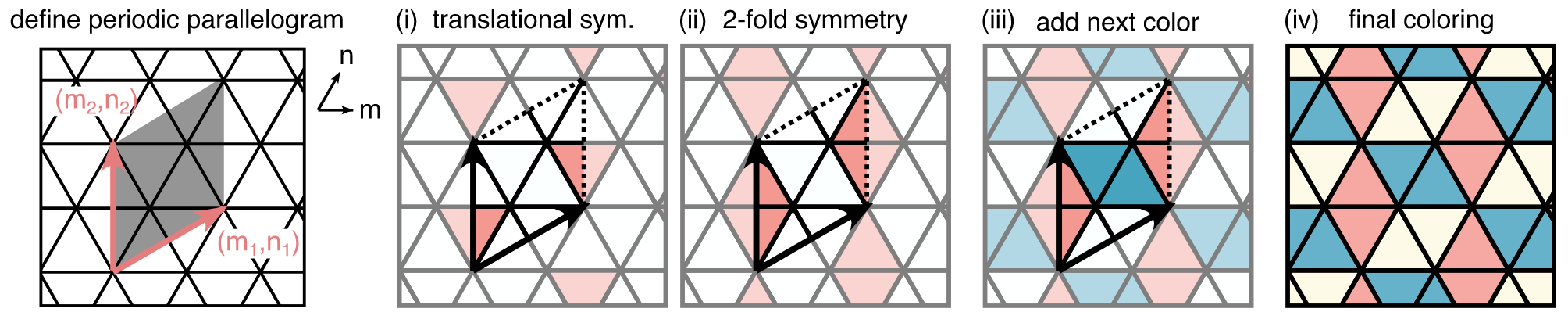}
    \caption{ \textbf{Generation of colorings.} Outline for coloring generation scheme. First, choose a parallelogram that will tile the plane, defined by primitive vectors ($m_1, n_1$) and ($m_2, n_2$). To color in the parallelogram, and subsequently the plane, we use the following steps: (i) assign a color to an arbitrary triangle in the parallelogram and color any other triangles that can be reached by any of the primitive vectors, (ii) color any triangles that can be reached by a colored triangle from a 2-fold rotation about the center of the parallelogram, (iii) repeat the previous steps for an uncolored triangle with a new color until (iv) all triangles in the parallelogram have a color assigned.}
    \label{Sfig:PatternGen}
\end{figure*}

Here we generate coloring patterns by prescribing a desired unit cell for the tiling and then using symmetry to color in the pattern. In Fig.~\ref{Sfig:PatternGen}, we outline the basic procedure. First, we choose a primitive cell for the coloring, which is a parallelogram that can tile the plane by translation and is defined by two primitive vectors for its side length. Next, we assign a color to an arbitrary triangle that lies within the parallelogram. Since we require that our parallelogram has both translational symmetry and 2-fold rotational symmetry, some triangles within the parallelogram are required to have the same color to preserve symmetry. After assigning a color, we color in any symmetric triangles in the pattern. We continue this method of coloring until all the triangles in the parallelogram have been assigned a color. A benefit of this type of pattern generation is that it is simple to create a multi-color assembly that targets a specific tubule state. Since the primitive vectors of the unit cell define the periodicity of the coloring, they also inform one of the tubule states that will be geometrically commensurate with that coloring.

\section{Design of interaction sequences} \label{SIsec:intDesign}

To have all of the required interactions for our multi-component assemblies, we need to have a library of unique interactions that have low crosstalk. We use an algorithm described by Nadrian Seeman~\cite{Seeman1990} to construct our library. A sketch of the algorithm is as follows. First, we choose a set of bases (letters) that we will use to construct our sequences, such as AGT for three letters. Using these letters, we create a list of all their permutations of up to some length $N$.  We call this set our dictionary. After populating the dictionary with all permutations, we remove all entries that contain the same letter at least three times in a row, such as AAA. This dictionary is then used to generate a new sequence (word) of some target length $M$, where $M$ is greater than $N$. To create a word,  (i) we pick an entry at random from our dictionary and set it to be the start of the word, (ii) we remove the entry from the dictionary as well as its complement, if it is there, (iii) we pick a new entry from our dictionary that starts with the last $N$-1 letters of our current word and add the last letter of our picked entry to the word, (iv) we remove the picked entry and its complement from the dictionary, (v) we repeat steps (iii) and (iv) until the word reaches the target length $M$, and (vi) we repeat steps (i-v) until the dictionary is depleted to the point that new $M$-letter words cannot be formed.

Using a four-letter alphabet (ACGT) with five-letter dictionary entries, we generated a list of 187 six-letter words. From this list, we choose 72 words such that they, and their compliments, had the lowest off-target interactions. These sequences are listed in Table~\ref{tab:seqlibrary} along with their calculated binding free energy to their complements at 1~M NaCl~\cite{SantaLucia2004}.

\section{Library of side interactions}

\begin{figure*}[!ht]
    \centering
    \includegraphics[width=\linewidth]{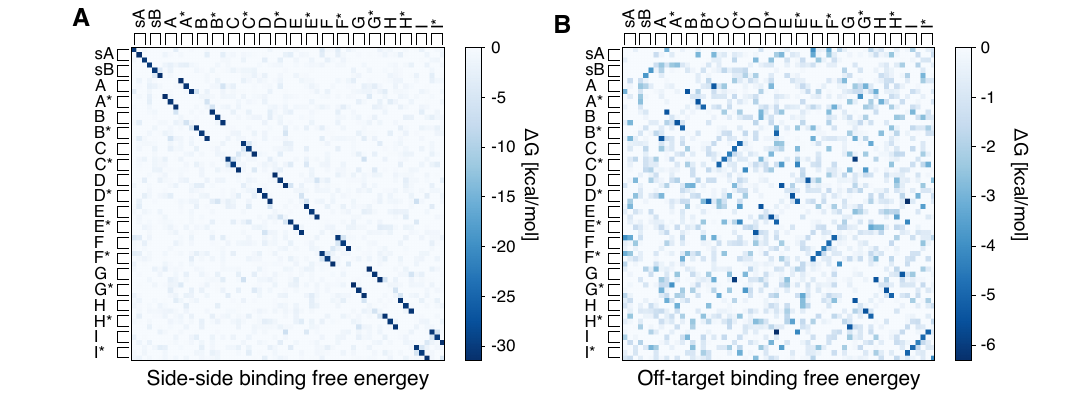}
    \caption{ \textbf{Side-side interaction energy.} (A) Free energy of binding between different sets of interactions is shown in Table~\ref{tab:sideinteractions}. Each set of three rows and columns (denoted by the bracket) corresponds to the side 1, 2, 3 sequences in that order. (B) To highlight the weaker binding energy of the off-target interactions, we set all intended binding interaction energies to 0 kcal/mol and rescaled the colorbar.}
    \label{Sfig:sidebindingenergy}
\end{figure*}

Given a list of sequences with minimal crosstalk, we also need to generate a library of interactions for each side of each subunit species. For any multi-component design, there can be both self-complementary and non-self-complementary interactions. For each triangular subunit, we want six strands on each of its three sides. Given this design criteria, we create a list of multiple self-complementary and non-self-complementary sequences for each side. These strands are listed in Table~\ref{tab:sideinteractions} along with the sum of their binding free energies. The set of strands for each side was chosen so that they would have similar total binding free energies to each other. Due to the limited size of our interaction sequence library, some sets of six strands share the same sequence. In these cases, the locations of the sequences were chosen to minimize off-target binding. Estimates of binding energies are shown in Fig.~\ref{Sfig:sidebindingenergy}.

\section{Analysis of tubule distributions}

After assembly experiments, we observe the assembled structures with TEM. We find many tubules that are deposited upon the mesh that present a variety of tubule types. Since our tubules are made of a discrete number of subunits, we know each tubule, or section of a tubule, can be identified with a unique ($m,n$) pair, where $m$ and $n$ describe the shortest path around the circumference of the tubule while moving allong the \textbf{m} and \textbf{n} lattice directions, as shown in Fig.~1B of the main text. As described in the work of Hayakawa et al.~\cite{Hayakawa2022}, each type of tubule can be identified with a specific circumference and maximum seam angle. By measuring each of these quantities for all tubules we image, we can create a distribution of the tubule states that are accessible to the system.

\begin{figure*}[!ht]
    \centering
    \includegraphics[width=\linewidth]{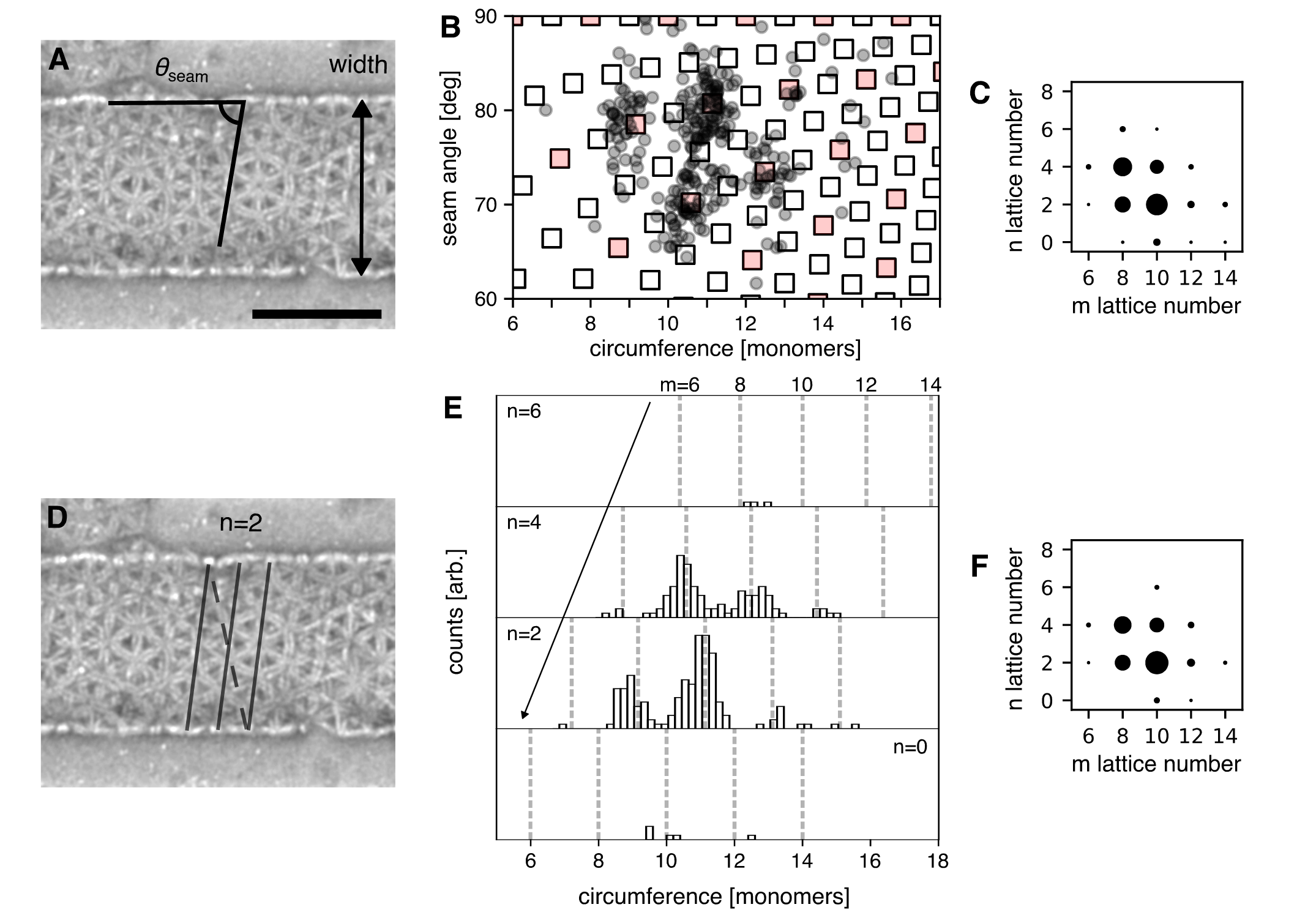}
    \caption{ \textbf{TEM tubule analysis} (A-C) Width and seam angle analysis. (A) For any tubule, one can measure the seam angle, $\theta_\mathrm{seam}$, and the width on the TEM gird. This image is from the 4-color experiment shown in Fig.~2 for the (10,0) monomer in the main text. Scale bar is 200 nm. (B) Plot of the seam angle versus the diameter of the tubule. We take the diameter to be twice the measured width. Open square denote all discrete tubule states, filled squares are the states allowed by the coloring of triangles. (C) By clustering points to the nearest allowed state from the seam angle versus width data, we construct the ($m,n$) distribution for the assembly. (D-F) Width and \textit{n} analysis. (D) By looking at the lattice seams of the tubule one can directly infer the $n$ state of the tubule. The solid line shown seams for one layer of the tubule and the dashed line is the seam on the other layer. In this image we see that there are exactly two rows of triangles that span the dashed line, so we know this is an $n=2$ tubule. (B) Histograms of the diameters of tubules for each corresponding $n$ number measured in TEM. Dashed vertical lines are the diameter for tubules with $m$ numbers 6 to 14 with the corresponding $n$ number and allowed by the coloring. Note that the histograms show peaks only at the allowed $m$ numbers. (F) By clustering points in each histogram to their nearest allowed $m$ number we construct the ($m,n$) distribution.}
    \label{SFig:TEMAnalysis}
\end{figure*}

We illustrate this analysis by looking at the data set for the 4-color isotropic case from Fig.~2E in the main text. In Fig.~\ref{SFig:TEMAnalysis}A, we show a TEM image of a tubule and the measured seam angle and width in the image. Measuring many tubules allows us to build up a distribution of seam angles and circumferences, Fig.~\ref{SFig:TEMAnalysis}B. We assume that the measured width is half the circumference since tubules become flattened during grid prep~\cite{Hayakawa2022}. Since the tubule does not lie perfectly flat, may crack slightly when deposited on the TEM grid, or have a varying amount of stain, there is some spread in the values away from those expected for discrete tubules. For the case of a multi-component assembly, only some of the states are expected to be allowed, as shown by the red squares. The data cluster about these points, but the spread is comparable to the separation between possible tubule types. Since the AuNP labeling and tomography experiments show high fidelity of interaction specificity, we cluster these points to their nearest allowed state to construct an ($m,n$) distribution for the assembly, Fig.~\ref{SFig:TEMAnalysis}C. The nearest state is the one that minimizes the sum in quadrature of the differences between the experimental and the expected values for discrete, flattened tubules of the seam angle and circumference.

To increase our confidence in the distributions we measure, we perform an additional analysis in which we identify the $n$ number directly from TEM images. In Fig.~\ref{SFig:TEMAnalysis}D, we show that the seams of triangles in both layers of the tubule can be seen. By matching up where the seams for two layers meet, we can directly infer the $n$ number for each tubule. This procedure is a more reliable method of determining the $n$ number than using the seam angle since those data show a broad spread compared to the separation between discrete tubule states. After determining $n$, we construct histograms of the measured circumferences for each set of unique $n$ numbers, as shown in Fig.~\ref{SFig:TEMAnalysis}E. Looking at the circumferences, we see distinct peaks that match up well with the states that are allowed from the coloring. Again we can cluster these points to the nearest allowed tubule state and construct an ($m,n$) distribution ( Fig.~\ref{SFig:TEMAnalysis}F). This method provides a quantitatively similar distribution to the one described above. For all distributions shown in the main text, we use this second method.

\section{Helfrich model of thermal fluctuations} \label{SIsec:Helfrich}

\subsection{Predicting tubule-type distributions}

To understand which neighboring tubule states are within the reach of thermal fluctuations, we consider how fluctuations of curvature for a growing sheet relate to the mechanical properties of the sheet. We use a Helfrich model for the elastic energy of the sheet~\cite{helfrich1986size,Helfrich1988}, which goes as $E = \frac{1}{2}BA(\Delta \kappa_\perp)^2$, where $A$ is the area of the assembly, $B$ is the bending rigidity, and $\Delta\kappa_\perp$ is the fluctuation of the curvature in the circumference direction. From prior work~\cite{Videbaek2022,Hayakawa2022,Fang2022}, we assume that the tubule type is frozen at the point at which the triangular sheet closes on itself to form a tubule. This assumption means that the dispersity of states in assembly can be directly tied back to the fluctuations of the pre-closure sheet. If the binding energies for each side of the triangular subunits are similar, we assume that the assembly grows as a patch-like disk. When enough curvature has accumulated so that it can close, the diameter of the disk should match the circumference of the closed tubule and have an area of $A=\pi^3 R^2$, where $R$ is the radius of curvature of the tubule. From this set of assumptions, we can estimate the fluctuation of the curvature as:
\begin{equation}
    (\Delta\kappa_\perp)^2 \approx \frac{1}{R^2}\left(\frac{\Delta R}{R}\right)^2.
\end{equation}
Noting that the circumference fluctuations are proportional to the radius fluctuations, we find that the Helfrich energy is 
\begin{equation}
    E_\mathrm{H}=\frac{1}{2}\pi^3B\left(\frac{\Delta C}{C}\right)^2.
\end{equation}
Assuming that the circumferences of the tubules follow the Boltzmann distribution, $P=\exp(-E_\mathrm{H}/\mathrm{k_B}T)/Z$, where $Z$ is the partition function, then the standard deviation of the circumference goes as $\Delta C \propto C/\sqrt{B}$.

We extend this Helfrich model to the ($m,n$) space of tubule types by considering a simple elastic model for the energy of our assembly. Since the triangular monomers have a preferred binding angle on each side,  any deviation from this angle will result in an elastic energy cost. For an assembly with $N$ components, the elastic energy can be estimated as 
\begin{equation}
    E=\frac{1}{4}NB\sum_{i\in1,2,3} (\theta_i - \theta_{0,i})^2,
    \label{SEqn:HelfrichModel}
\end{equation}
where $\theta_{0,i}$ is the preferred binding angle and $\theta_i$ is the actual binding angle for side $i$. We can estimate the size of an assembly at closure as we did above, finding $N=(4\pi/\sqrt{3})(C/l_0)^2$, where $l_0$ is the edge length of a subunit. As a note, we can estimate the difference in binding angles for different size tubules and find that $\Delta \theta \propto \Delta C/C^2$. Putting this result into Eqn.~3 gives the same scaling as in Eqn.~2. Since both $N$ and $\theta_{0,i}$ depend upon $m$ and $n$, this model gives a state-dependent probability $P(m,n)$ that we also assume follows a Boltzmann distribution.

\subsection{Scaling law for isotropic colorings}

\begin{figure*}[!ht]
    \centering
    \includegraphics[width=\linewidth]{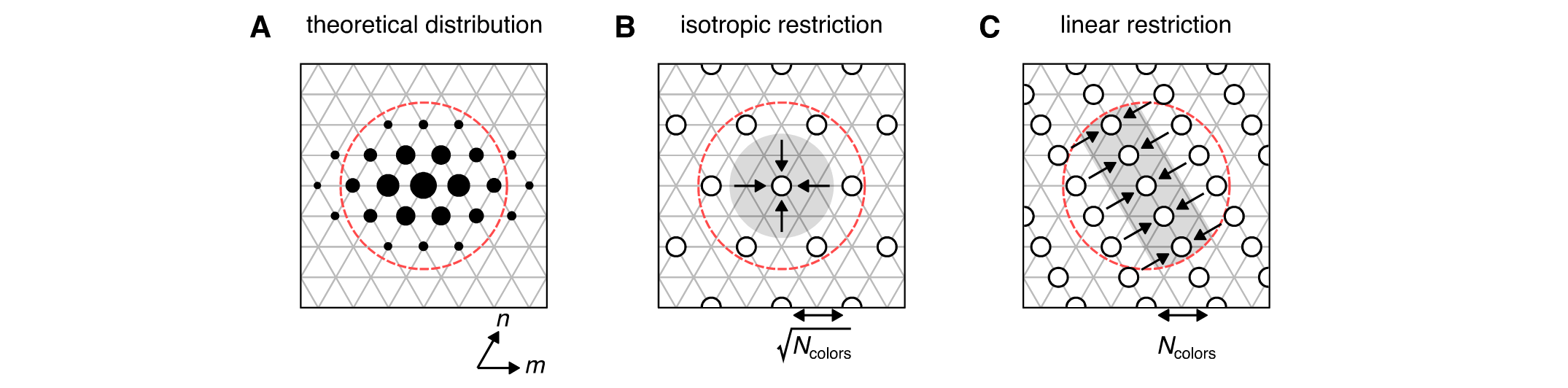}
    \caption{ \textbf{Scaling for selectivity with $N_\mathrm{colors}$.} (A) Tubule distribution generated from Eqn.~3 targeting a (10,0) state with a bending modulus of 8~\kT. (B) By using isotropic colorings we remove possible states from around the target state in an area. The distance between states scales with $\sqrt{N_\mathrm{colors}}$. Since some states can no longer close into the tubules due to the coloring pattern, probability from these states gets funneled into neighboring available states. (C) With linear colorings the separation between states grows as $N_\mathrm{colors}$. White circles on the plane denote similar vertices based on an unspecified coloring.}
    \label{Sfig:HelfrichEng}
\end{figure*}

Making some simplifying assumptions about this model lets us infer a scaling law for how the increase in selectivity of tubule states depends on the number of colors and properties of the tubule. In Fig.~\ref{Sfig:HelfrichEng}A, we show an ($m,n$) distribution of tubule states generated by Eqn.~3 overlayed on the vertices of the triangular lattice of monomers. The distribution of states has two well-defined axes, one shorter than the other, and the distribution of these states is fairly Gaussian, owing to the $\theta^2$ dependence of the elastic energy. To account for the coloring patterns of states, we note that for isotropic patterns, similar vertices in the coloring are separated by a distance of $\sqrt{N_\mathrm{colors}}$. When this coloring pattern is imposed on the tubule some states become inaccessible and their probabilities are funneled into neighboring available states (Fig.~\ref{Sfig:HelfrichEng}B). To get an estimate for the increase in selectivity, we integrate the tubule state probability distribution within a circle of radius $\sqrt{N_\mathrm{colors}}$, whose origin is the center of the distribution. To make this calculation simpler, we assume that the tubule distribution is radially symmetric such that the selectivity takes the form
\begin{equation}
    P \propto \int_0^{2\pi}\int_0^{\sqrt{N_\mathrm{colors}}} r\exp(-r^2/(C^2/B))\mathrm{d}r\mathrm{d}\theta,
\end{equation}
where the standard deviation of the Gaussian is taken to be $C/\sqrt{B}$. Evaluating this integral gives an approximate equation for the selectivity of
\begin{equation}
    P \propto \exp(-N_\mathrm{colors}B/C^2).
\end{equation}
The important takeaway from this result is that the selectivity depends upon $N_\mathrm{colors}B/C^2$, which we use to rescale our experimental data in Fig.~3 in the main text. Though we make approximations to make this calculation straightforward, numerical studies of Eqn.~3 find the same scaling for the selectivity~\cite{Videbaek2022}.

\subsection{Scaling law for linear colorings}

Using the same type of argument outlined above, we can make an estimate for the scaling law for the selectivity of the width- and pitch-control experiments outlined in Fig.~4 in the main text. As shown in Fig.~\ref{Sfig:HelfrichEng}C, the linear colorings have states that are funneled to a seam of allowed points. To get an estimate of the scaling for how these colorings change selectivity, we need only integrate over one dimension of the Gaussian distribution. Since the separation between the seams grows as $N_\mathrm{colors}$, we have that
\begin{equation}
    P \propto \int_{-\infty}^{\infty} \int_{-N_\mathrm{colors}}^{N_\mathrm{colors}} \exp(-(x^2 + y^2)/(C^2/B))\mathrm{d}x \mathrm{d}y,
\end{equation}
where $x$ and $y$ correspond to a change in basis with $x$ pointed along the seam of similar states and $y$ perpendicular to it. This integration results in
\begin{equation}
    P\propto \mathrm{erf}(N_\mathrm{colors}\sqrt{B}/C).
\end{equation}
This equation reveals that the scaling for the selectivity has changed to depend upon $N_\mathrm{colors}\sqrt{B}/C$, which we use to rescale the data in Fig.~4E.

\subsection{Comparison with experiment}

\begin{figure*}[!ht]
    \centering
    \includegraphics[width=\linewidth]{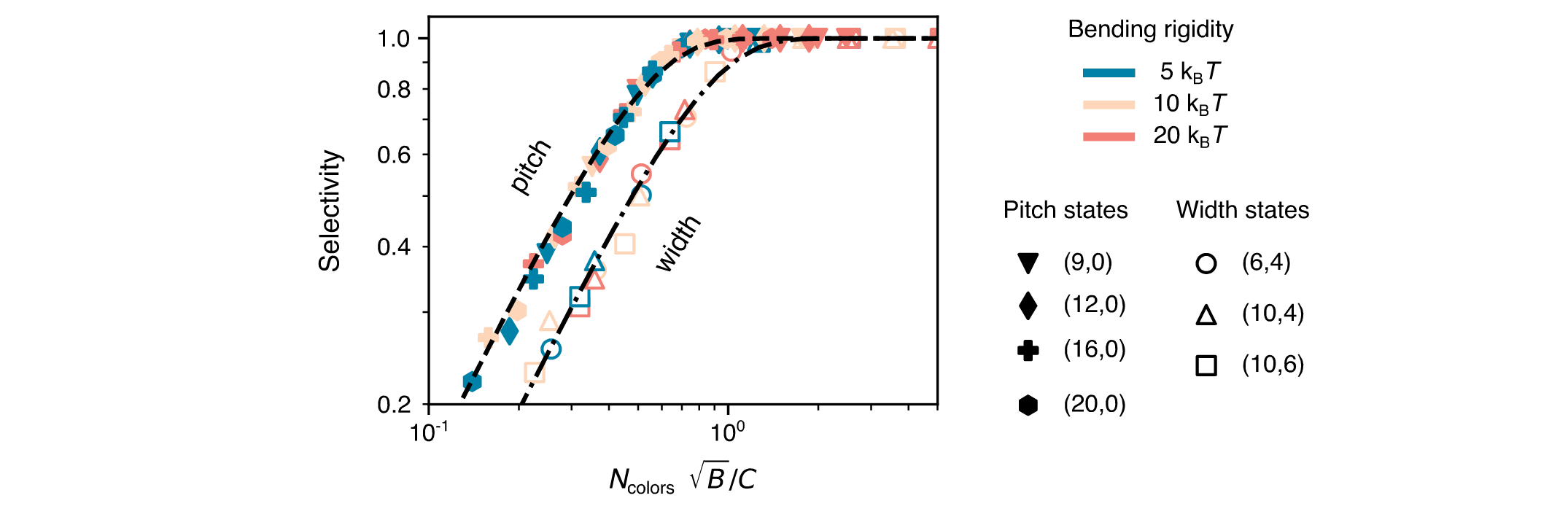}
    \caption{ \textbf{Numerical simulations for pitch and width controlled tubules.} Selectivity versus the rescaled linear number of colors for pitch- and width-controlled tubules based on the Helfrich model. Open symbols are width-controlled and full symbols are pitch-controlled. A variety of bending rigidities and tubule circumferences are shown. The dark lines show approximations to the curves and are used in Fig.~4E in the main text.}
    \label{Sfig:HelfrichSim}
\end{figure*}

To compare this model with experiment, we look at distributions of tubules generated by Eqn.~\ref{SEqn:HelfrichModel} and see how the distributions change when different tubule colorings are used to restrict states. For the isotropic colorings, we use the results from Videb{\ae}k et al.~\cite{Videbaek2022}. For the pitch- and width-controlled cases, we generate distributions for a variety of bending rigidities and preferred tubule geometries, and restrict their distributions using linear colorings. Scaling this data with $N_\mathrm{colors}\sqrt{B}/C$ collapses the data and shows that the pitch- and width-controlled cases split into two curves, as seen in Fig.~\ref{Sfig:HelfrichSim}. These two curves arise due to the slight asymmetry of the tubule distribution in the ($m,n$) space (Fig.~\ref{Sfig:HelfrichEng}A). The dashed lines in Fig.~\ref{Sfig:HelfrichSim} show the same curves as shown in Fig.~4E in the the main text.

\section{Length-controlled tubule gel}

\begin{figure*}[!ht]
    \centering
    \includegraphics{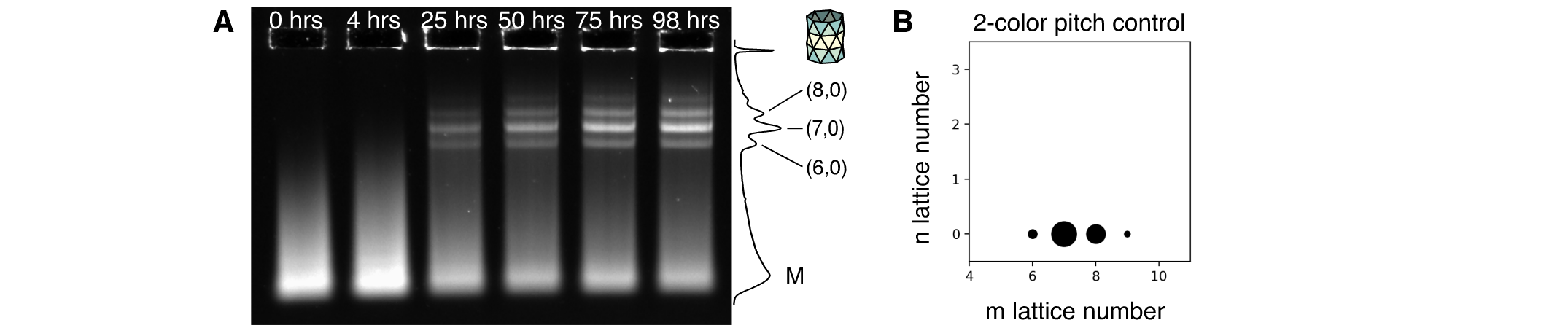}
    \caption{ \textbf{Kinetics of length controlled tubules} (A) Scan from a gel electrophoresis experiment of the 3-color length controlled tubules of the (6,0) monomer for different assembly times. On the right is a plot of the fluorescence intensity for the 98 hr assembly. M denotes the monomer band. (B) Tubule distribution for a 2-color pitch-controlled assembly using the (6,0) monomer.}
    \label{Sfig:TubeletGel}
\end{figure*}

Since the length-controlled tubules are finite assemblies, it is possible to observe their evolution toward equilibrium with gel electrophoresis. We prepare assemblies at different points in time and run them in a 0.5\% agarose gel to separate different-sized assemblies. Figure~\ref{Sfig:TubeletGel}A shows a representative gel of this type of kinetics experiment. At early times, we see that the assemblies are mostly monomers with some signal coming from larger assemblies, though it is possible that some of this occurs from assembly at low temperatures in the pocket of the gel. At later times, we see the emergence of distinct bands. We infer the geometries of the tubules in these bands by comparing them to the ($m,n$) distribution of a 2-color pitch-controlled tubule experiment (Fig.~\ref{Sfig:TubeletGel}B). In the tubule distribution, the smallest structure is a (6,0) tubule, the most probable state is (7,0), and there is a tail at larger \textit{m}. We see a similar structure for the bands in the gel scan and conclude that larger diameter length-controlled tubules travel slower in the gel. We note that we did not characterize the ($m,n$) distribution from TEM images of length-controlled tubules since the length-controlled tubules often break during deposition on the grid, making interpretation of widths unreliable for these small structures.

\section{Kinetics of tubule growth}

\begin{figure*}[!ht]
    \centering
    \includegraphics[width=\linewidth]{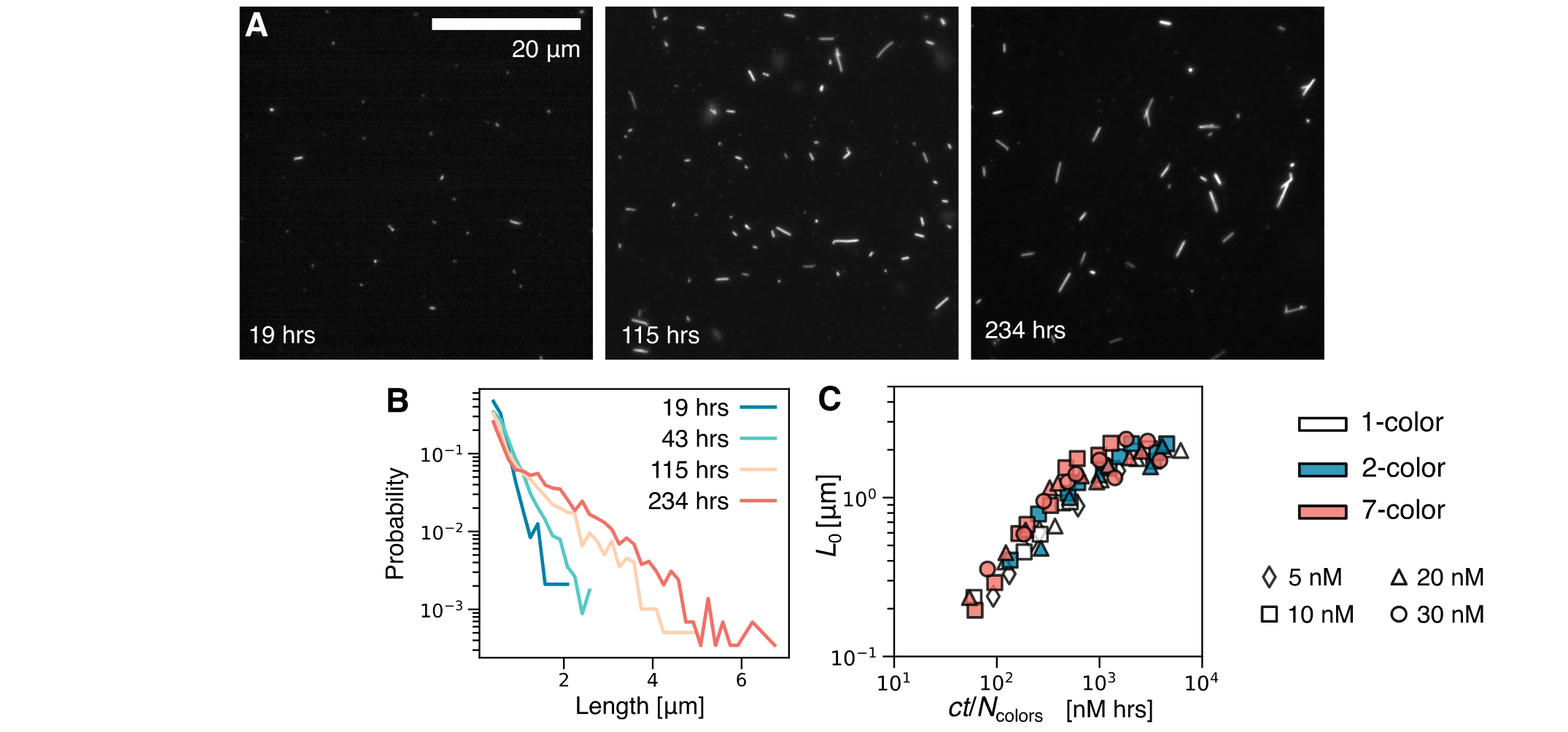}
    \caption{ \textbf{Kinetics of multi-color tubule growth} (A) Fluorescent microscopy images of a 7-color tubule assembly at different points of time. (B) Length distributions of tubules at different points in time. (C) Average length of tubules, $L_0$, plotted against a rescaled time, $ct/N_\mathrm{colors}$.}
    \label{Sfig:Kinetics}
\end{figure*}

To see how changing the number of components in an assembly changes the growth rate, we perform kinetics experiments that quantifies the length of tubules over time. Since our assemblies grow to several microns in length, it is simple to observe them with epi-fluorescence microscopy. After staining the assemblies with an intercalating dye (YOYO-1), we take images of the tubules at different points in time. Figure~\ref{Sfig:Kinetics}A shows exemplary images for a 7-color experiment over ten days, with larger structures appearing at later times. To quantify the change in the lengths, we binarize the images, fit the particles to ellipses, and take the fitted major lengths as the lengths of the tubules. By performing this analysis for a set of thirty images, we generate a distribution of lengths for different points in time (Fig.~\ref{Sfig:Kinetics}B). For each time point, we extract the mean length, $L_0$, of the distribution. Simple models for the growth of one-dimensional filaments suggest that the relevant parameters for rescaling the assembly time scale are the initial monomer concentration, $c$, and the number of colors, $N_\mathrm{colors}$, since these directly impact the rate of monomer addition for a growing assembly~\cite{Phillips2012physical}. In Fig.~\ref{Sfig:Kinetics}C, we plot $L_0$ versus a rescaled time, $ct/N_\mathrm{colors}$, and find that our data collapse to a single curve for a range of $N_\mathrm{colors}$ and initial concentrations.

\subsection{Analysis of fluorescence images for kinetics}

Due to the way that we binarize our images, there is a lower bound for the size of objects that we can detect. This lower bound directly manifests in the distributions that we measure. Since we want to measure the average length, $L_0$, of the tubule distribution, we need to incorporate this lower bound into extracting the correct average length. We implement this correction by integrating over an exponential distribution from this lower bound, $L_c$, instead of 0, 
\begin{equation}
    \frac{\langle L^2 \rangle}{\langle L \rangle} = \frac{\int_{L_c}^\infty (l^2/L_0) \exp(-l/L_0) \mathrm{d}l }{\int_{L_c}^\infty (l/L_0)
 \exp(-l/L_0)\mathrm{d}l  } = \frac{2L_0^2 + 2L_0L_c +L_c^2}{(L_c+L_0)},
\label{SEqn:HistLength}
\end{equation}
where $L_c$ is the cutoff length scale. By measuring the experimental value of $\langle L^2 \rangle/\langle L \rangle$ and using Eq.~~\ref{SEqn:HistLength}, we can solve for $L_0$ of the tubule lengths distribution. For all measured distributions, we use a consistent cut-off of $L_c=500$ nm.

\section{Trade-offs in selectivity}

\begin{figure*}[!ht]
    \centering
    \includegraphics[width=\linewidth]{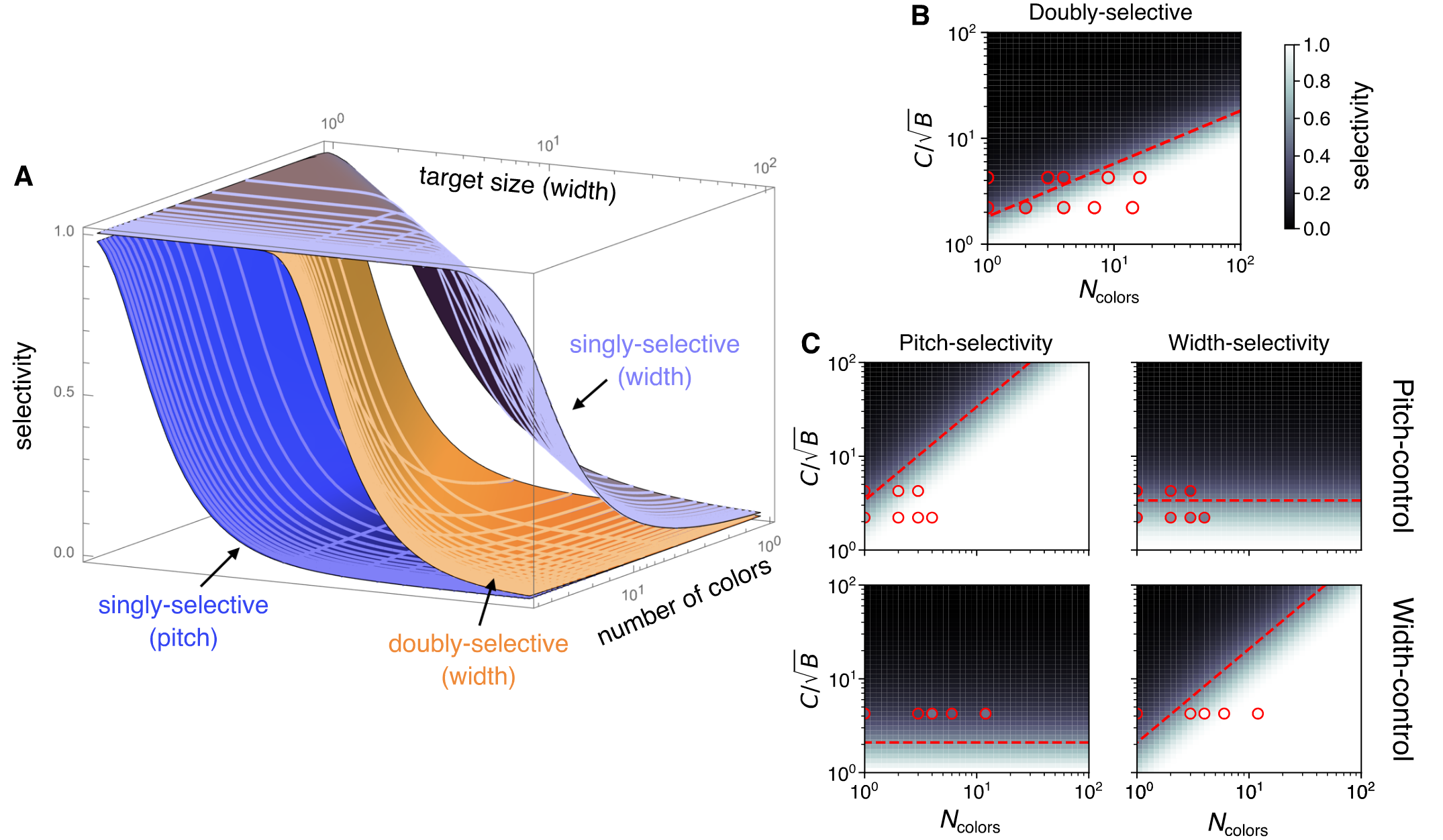}
    \caption{ \textbf{Selectivity in tubules for different strategies.} (A) Surface plot of selectivity of the tubule width plotted against the target tubule size and the number of colors. The orange surface corresponds to the scheme of targeting a single tubule state with an isotropic coloring pattern (called doubly-selective), as shown in the main text Fig. 2 and 3. For the width-control scheme (singly-selective), as shown in the main text Fig. 4, the light blue surface shows the selectivity of the tubule width while the dark blue shows the corresponding pitch selectivity. (B) Shows a color map of the doubly-selective strategy. Circle points show the experimental data shown from Fig.~3 and share the same colorbar as the theoretical points. (C) shows the pitch-  (left) and width-selectivity (right) for both the pitch-control (top) and width-control (bottom) schemes. This demonstrates the trade-off that though the scaling of selectivity with the self-limited length is better while constraining just a single dimension of the assembly, one loses any improvement of selectivity in the other dimension of the assembly.}
    \label{Sfig:TradeOff}
\end{figure*}

As described in the main text, we introduced a strategy for targeting either the pitch or the width of a tubule to get a more favorable scaling for the complexity by limiting a single dimension of the thermal fluctuations. This allows one to prescribe a specific assembly property with significantly fewer colors than when selecting for a single assembly state. Of course, this comes with the trade-off that the unconstrained dimension gains no benefit in selectivity with increasing complexity. We illustrate this in Fig.~\ref{Sfig:TradeOff} by plotting the selectivity of assemblies using either the singly- or doubly-selective strategies. Based on our experimental data, Fig.~3 and 4 of the main text, and numerical calculations, Fig.~\ref{Sfig:HelfrichSim}, the distributions of tubules match our Helfrich model. Using this model we plot the theoretical width-selectivity for assemblies with either the singly- or doubly-selective strategies against both the assembly size and the number of colors, Fig.~\ref{Sfig:TradeOff}A. For another way to view this trade-off, we also plot separately the doubly-selective prediction, Fig.~\ref{Sfig:TradeOff}B, and singly-selective prediction, Fig.~\ref{Sfig:TradeOff}C, along with our experimental data. For the singly-selective plots, we show both the pitch- and the width-selectivity. These color maps emphasize the improvement of selectivity for larger self-limiting sizes, 

\clearpage

\begin{longtable}{p{0.13\linewidth} p{0.13\linewidth} 
 p{0.15\linewidth} p{0.05\linewidth}p{0.13\linewidth} p{0.13\linewidth} p{0.15\linewidth}}
\caption{\textbf{Interaction sequence library.} A list of 72 unique sequences generated using the algorithm described in Suppl. Sec.~\ref{SIsec:intDesign}, along with the complementary sequence and an estimate of their binding free energy.} \label{tab:seqlibrary} \\
\input{SequenceLibrary}

\end{longtable}

\clearpage

\begin{longtable}{p{0.11\linewidth} p{0.11\linewidth} p{0.11\linewidth} p{0.11\linewidth}p{0.11\linewidth} p{0.11\linewidth} p{0.11\linewidth} p{0.14\linewidth}}
\caption{\textbf{Side interactions for multicomponent assemblies.} A list of the set of six interaction sequences that make up a side interaction of a monomer and an estimate of their summed binding free energy. The first six sets of sequences are self-complimentary, e.g. Position 1 binds to Position 6, Position 2 binds to Position 5, and Position 3 binds to Position 4. For the rest of the sets, X is complimentary to X*.} \label{tab:sideinteractions} \\
\input{MultispeciesSideInteractions}
\end{longtable}


\begin{longtable}{p{0.09\linewidth} p{0.09\linewidth} p{0.07\linewidth} p{0.07\linewidth}p{0.07\linewidth} p{0.05\linewidth} p{0.09\linewidth} p{0.09\linewidth} p{0.07\linewidth} p{0.07\linewidth}p{0.07\linewidth} }
\caption{\textbf{Subunit interactions for isotropic colorings.} This table enumerates the different side interactions from Table~\ref{tab:sideinteractions} used to generate the colorings used in Fig.~2 and 3 in the main text for the (6,0) tubule.} \label{tab:isopatinteractions-6_0} \\
\input{IsotropicPatternInteractions-6-0}
\end{longtable}

\begin{longtable}{p{0.09\linewidth} p{0.09\linewidth} p{0.07\linewidth} p{0.07\linewidth}p{0.07\linewidth} p{0.05\linewidth} p{0.09\linewidth} p{0.09\linewidth} p{0.07\linewidth} p{0.07\linewidth}p{0.07\linewidth} }
\caption{\textbf{Subunit interactions for isotropic colorings.} This table enumerates the different side interactions from Table~\ref{tab:sideinteractions} used to generate the colorings used in Fig.~2 and 3 in the main text for the (10,0) tubule.} \label{tab:isopatinteractions-10_0} \\
\input{IsotropicPatternInteractions-10-0}

\end{longtable}

\begin{longtable}{p{0.18\linewidth} p{0.07\linewidth} p{0.07\linewidth}p{0.07\linewidth} p{0.05\linewidth} p{0.18\linewidth} p{0.07\linewidth} p{0.07\linewidth}p{0.07\linewidth} }
\caption{\textbf{Subunit interactions for linear tubule colorings.} This table enumerates the different side interactions from Table~\ref{tab:sideinteractions} used to generate the colorings used in Fig.~4 and 5 in the main text. For the pitch- and width-controlled tilings, we use the ``Linear tiling monomers" shown in the table below. These monomers have a repetitive nature to their interactions and the linear coloring can be terminated at any number of colors in the tiling. If the number of colors is even, we replace the final monomer with the ``Even periodic" monomer, where X* matches the interaction of the replaced monomer, i.e. if we made an 8-color linear tiling then we would use the ``Linear tiling monomers" 1 to 7 and an ``Even periodic" monomer with X* being D*, following the subunit ID 8 pattern. If the number of colors is odd, we replace the final monomer with the ``Odd periodic" interactions, swapping X* to the interaction of the replaced monomer. Different permutations of the side interactions can orient the linear colorings along different lattice directions. Pitch-controlled tubules have side identifications as shown in the table, while width-controlled tubules swap the Side 2 and Side 3 interactions. Similarly, to make length-controlled tubules we can take a linear tiling and make the final monomer a ``Capping monomer", that has no interaction on its side 3.} \label{tab:tubletinteractions} \\
\input{LengthConfinedTublets}
\end{longtable}

\clearpage


\begin{figure*}[!ht]
    \centering
    \includegraphics[width=\linewidth]{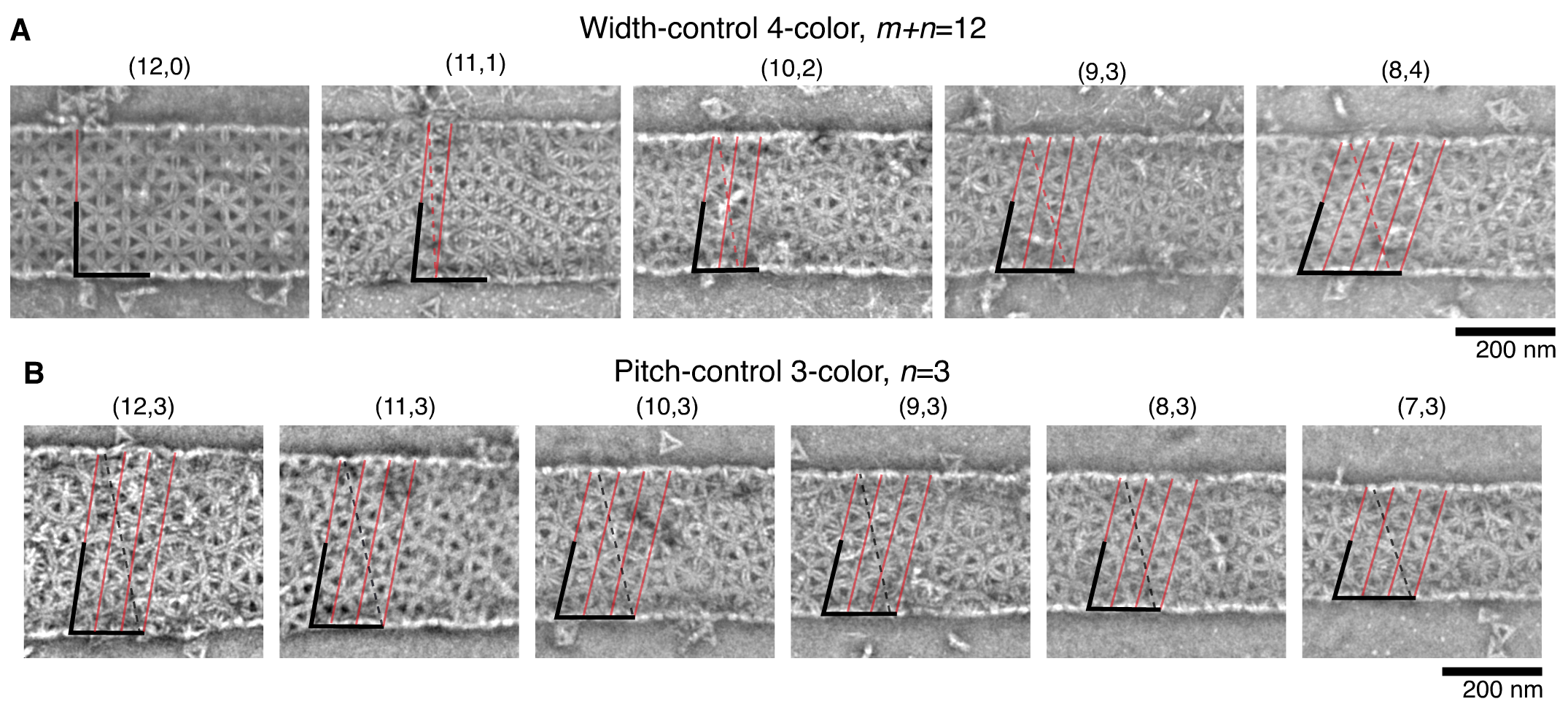}
    \caption{ \textbf{Gallery of width and pitch controlled tubules} (A) Tubules with $m+n=12$ from a 4-color width-controlled experiment. Tubules have nearly constant width, but varying pitch. (B) Tubules with $n=3$ from a 3-color pitch-controlled experiment. Tubules have constant pitch, but varying width. Both sets of experiments come from the distributions shown in Fig.~4A in the main text.}
    \label{Sfig:Gallery}
\end{figure*}

\begin{figure*}[!ht]
    \centering
    \includegraphics[width=\linewidth]{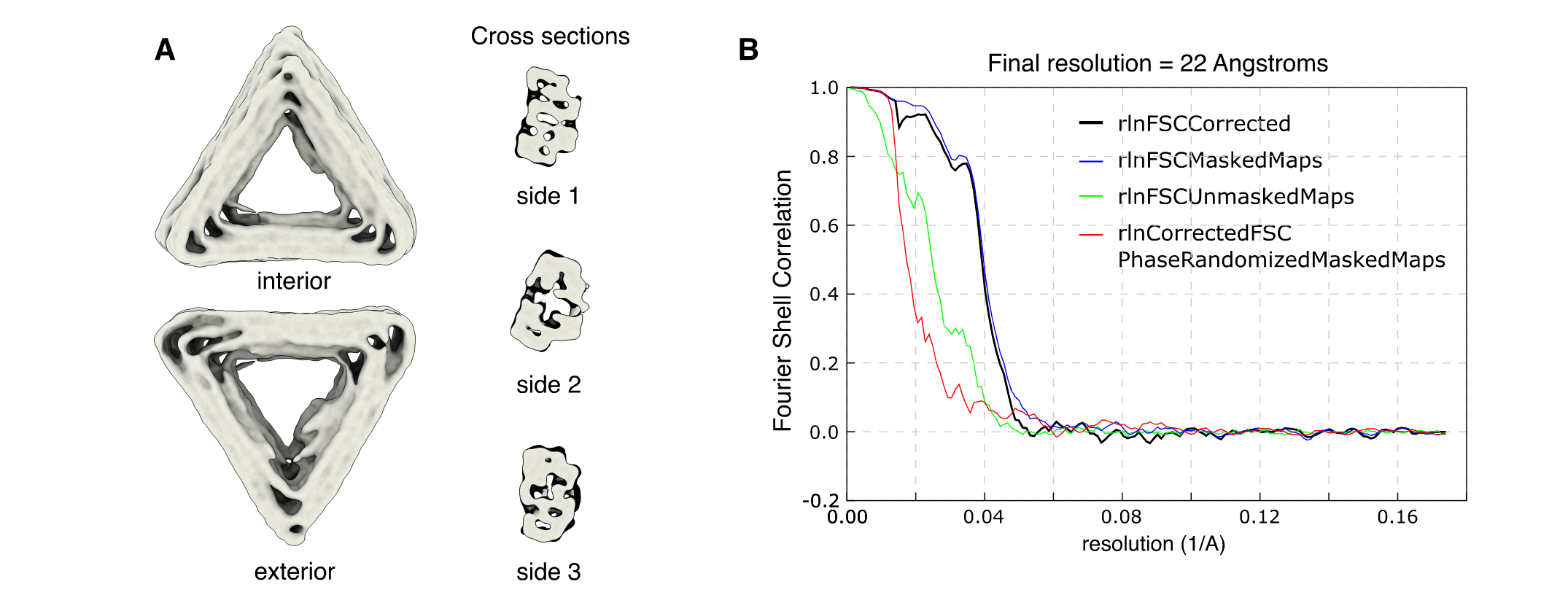}
    \caption{ \textbf{Cryo-EM reconstruction of (6,0) monomer.} (A) Views of the DNA origami monomer from the interior or exterior with respect to the tubules that form. Cross-sectional slices of the middle of each side. (B) Plot of the FSC curves used to estimate the resolution of the monomer reconstruction.}
    \label{Sfig:monomerCryo-6-0}
\end{figure*}

\begin{figure*}[!ht]
    \centering
    \includegraphics[width=\linewidth]{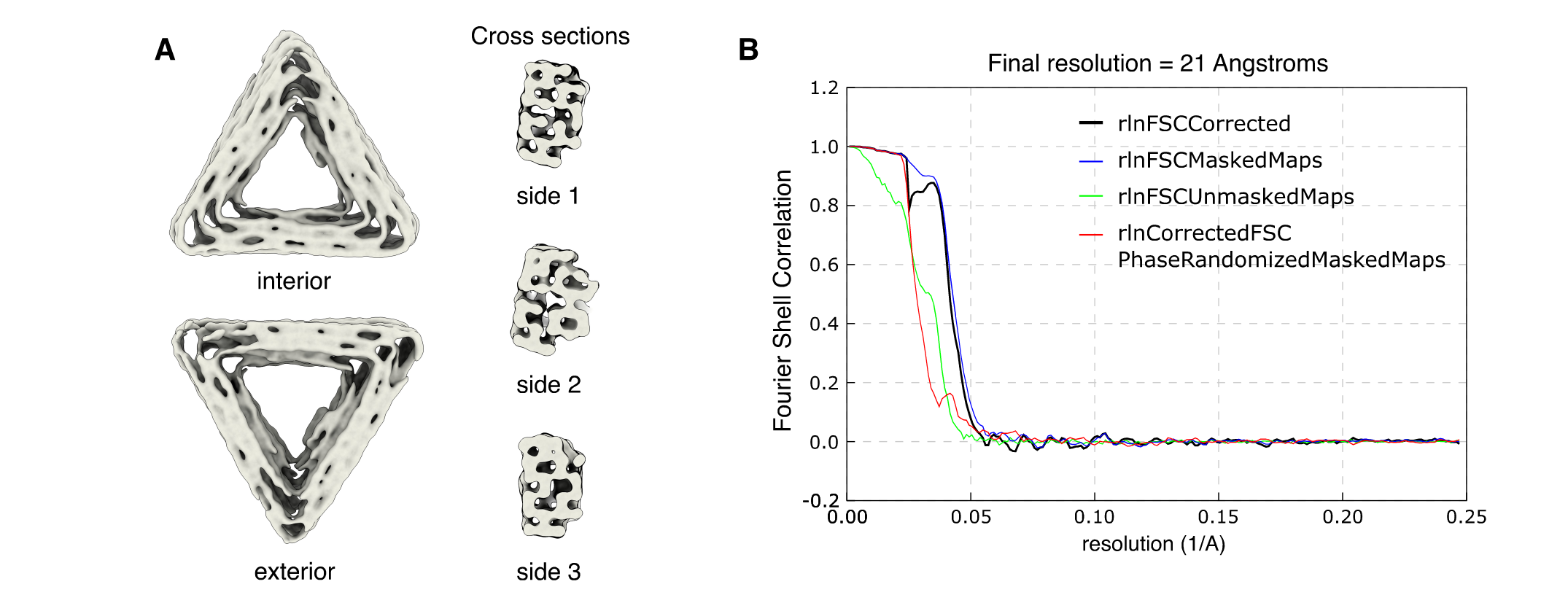}
    \caption{ \textbf{Cryo-EM reconstruction of (10,0) monomer.} (A) Views of the DNA origami monomer from the interior or exterior with respect to the tubules that form. Cross-sectional slices of the middle of each side. (B) Plot of the FSC curves used to estimate the resolution of the monomer reconstruction.}
    \label{Sfig:monomerCryo-10-0}
\end{figure*}

\begin{figure*}[!ht]
    \centering
    \includegraphics[width=\linewidth]{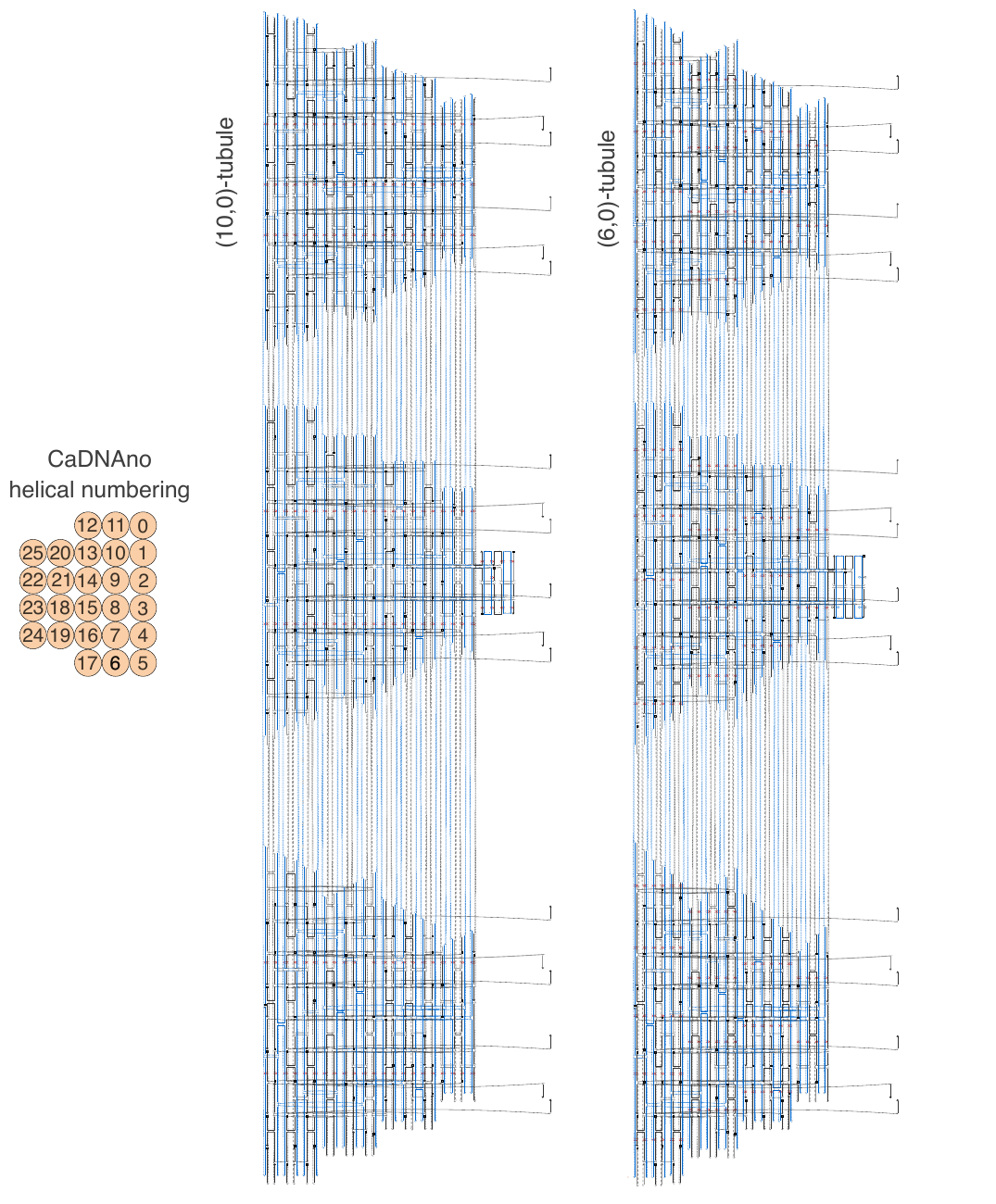}
    \caption{ \textbf{caDNAno design of DNA origami monomer with helical numbering.}}
    \label{Sfig:monomerCADNANO}
\end{figure*}

%% file: SequenceLibrary.tex
Sequence & Compliment & $\Delta$G [kcal/mol] & & Sequence & Compliment & $\Delta$G [kcal/mol] \\
\hline
\endfirsthead

Sequence & Compliment & $\Delta$G [kcal/mol] & & Sequence & Compliment & $\Delta$G [kcal/mol] \\
\hline
\endhead

\endfoot
\endlastfoot
ACTAGC	&	GCTAGT	&	-6.09	&	&	AGTTCC	&	GGAACT	&	-6.13	\\
AGTTAC	&	GTAACT	&	-5.01	&	&	CGATTA	&	TAATCG	&	-4.42	\\
TAGTCT	&	AGACTA	&	-4.53	&	&	ATTCTG	&	CAGAAT	&	-5.01	\\
CGATGG	&	CCATCG	&	-6.61	&	&	ATTCAG	&	CTGAAT	&	-5.01	\\
CCATTC	&	GAATGG	&	-5.61	&	&	CTTGAG	&	CTCAAG	&	-5.28	\\
CTTGGT	&	ACCAAG	&	-5.92	&	&	GTAGAT	&	ATCTAC	&	-4.42	\\
ATGCAC	&	GTGCAT	&	-6.73	&	&	GGATAA	&	TTATCC	&	-4.12	\\
TCGACA	&	TGTCGA	&	-5.67	&	&	TCATCC	&	GGATGA	&	-5.43	\\
TTGGAT	&	ATCCAA	&	-4.9	&	&	GGTATT	&	AATACC	&	-4.68	\\
TCAGAC	&	GTCTGA	&	-5.43	&	&	GGTAAT	&	ATTACC	&	-4.68	\\
GTCTAG	&	CTAGAC	&	-4.88	&	&	ACTGAG	&	CTCAGT	&	-5.85	\\
TACCTT	&	AAGGTA	&	-4.79	&	&	AGAGAT	&	ATCTCT	&	-5.08	\\
AGTCAG	&	CTGACT	&	-5.85	&	&	AGATAG	&	CTATCT	&	-4.42	\\
CTCGAA	&	TTCGAG	&	-5.54	&	&	TTCCTG	&	CAGGAA	&	-5.36	\\
TCCTTC	&	GAAGGA	&	-5.38	&	&	TTCCAT	&	ATGGAA	&	-4.9	\\
GATCTT	&	AAGATC	&	-4.7	&	&	GATATG	&	CATATC	&	-4.09	\\
CTGATC	&	GATCAG	&	-5.35	&	&	TTAACC	&	GGTTAA	&	-4.52	\\
TCCACA	&	TGTGGA	&	-5.49	&	&	AACATT	&	AATGTT	&	-4.81	\\
CAATAG	&	CTATTG	&	-4.16	&	&	TTGGCA	&	TGCCAA	&	-5.99	\\
TGATTG	&	CAATCA	&	-4.57	&	&	GACCTC	&	GAGGTC	&	-6.33	\\
CTAGGA	&	TCCTAG	&	-4.77	&	&	CCTATG	&	CATAGG	&	-5	\\
CACATC	&	GATGTG	&	-5.66	&	&	CTTAGG	&	CCTAAG	&	-4.95	\\
ACGAAG	&	CTTCGT	&	-6.29	&	&	TAACAG	&	CTGTTA	&	-4.24	\\
ACCTGA	&	TCAGGT	&	-5.93	&	&	TCTTCT	&	AGAAGA	&	-4.59	\\
ATGACA	&	TGTCAT	&	-5.14	&	&	GTACAT	&	ATGTAC	&	-4.73	\\
TACAGG	&	CCTGTA	&	-5.08	&	&	ATAAGT	&	ACTTAT	&	-4.22	\\
AACCTA	&	TAGGTT	&	-4.76	&	&	TTCAAT	&	ATTGAA	&	-4.06	\\
GAGACA	&	TGTCTC	&	-5.29	&	&	CTTACT	&	AGTAAG	&	-4.49	\\
GACAGA	&	TCTGTC	&	-5.29	&	&	AGTATC	&	GATACT	&	-4.75	\\
CGTCCA	&	TGGACG	&	-6.69	&	&	GTATGT	&	ACATAC	&	-4.73	\\
GCATCT	&	AGATGC	&	-6.09	&	&	ACAATT	&	AATTGT	&	-4.81	\\
TATTCC	&	GGAATA	&	-4.26	&	&	ACTAAC	&	GTTAGT	&	-5.01	\\
AGATTC	&	GAATCT	&	-5.03	&	&	CTTGTA	&	TACAAG	&	-4.24	\\
TTCTCA	&	TGAGAA	&	-4.34	&	&	CTACAC	&	GTGTAG	&	-5.33	\\
CTGTGA	&	TCACAG	&	-5.41	&	&	AAGTAG	&	CTACTT	&	-4.68	\\
TCGTAC	&	GTACGA	&	-5.59	&	&	AACTAT	&	ATAGTT	&	-4.22	\\

%% file: MultispeciesSideInteractions.tex
Name	&	Position 1	& 	Position 2	&	Position 3	&	Position 4	&	Position 5	&	Position 6	&	$\Delta$G [kcal/mol]	\\
\hline
\endfirsthead
Name	&	Position 1	& 	Position 2	&	Position 3	&	Position 4	&	Position 5	&	Position 6	&	$\Delta$G [kcal/mol]	\\
\hline
\endhead
sA side 1	&	ACTAGC	& 	AGTTAC	&	TAGTCT	&	AGACTA	&	GTAACT	&	GCTAGT	&	-30.57	\\
sA side 2	&	TTCAAT	& 	CCATTC	&	CTTGGT	&	ACCAAG	&	GAATGG	&	ATTGAA	&	-31.42	\\
sA side 3	&	TTAACC	& 	TCGACA	&	TTGGAT	&	ATCCAA	&	TGTCGA	&	GGTTAA	&	-30.23	\\
sB side 1	&	TCAGAC	& 	GTCTAG	&	TACCTT	&	AAGGTA	&	CTAGAC	&	GTCTGA	&	-30.17	\\
sB side 2	&	ATAAGT	& 	CTCGAA	&	TCCTTC	&	GAAGGA	&	TTCGAG	&	ACTTAT	&	-30.14	\\
sB side 3	&	GATCTT	& 	CTGATC	&	TCCACA	&	TGTGGA	&	GATCAG	&	AAGATC	&	-31.27	\\
A side 1	&	CAATAG	& 	TGATTG	&	CTAGGA	&	CACATC	&	ACGAAG	&	ACCTGA	&	-31.38	\\
A side 2	&	ATGACA	& 	TACAGG	&	AACCTA	&	GAGACA	&	GACAGA	&	ACTAAC	&	-30.57	\\
A side 3	&	GTACAT	& 	AGTCAG	&	CGATGG	&	CTTACT	&	AGTATC	&	GTATGT	&	-31.16	\\
A* side 1	&	TCAGGT	& 	CTTCGT	&	GATGTG	&	TCCTAG	&	CAATCA	&	CTATTG	&	-31.38	\\
A* side 2	&	GTTAGT	& 	TCTGTC	&	TGTCTC	&	TAGGTT	&	CCTGTA	&	TGTCAT	&	-30.57	\\
A* side 3	&	ACATAC	& 	GATACT	&	AGTAAG	&	CCATCG	&	CTGACT	&	ATGTAC	&	-31.16	\\
B side 1	&	GCATCT	& 	TATTCC	&	AGATTC	&	TTCTCA	&	CTGTGA	&	TCGTAC	&	-30.72	\\
B side 2	&	AGATAG	& 	TTCCTG	&	TTCCAT	&	GATATG	&	ATGCAC	&	AACATT	&	-30.31	\\
B side 3	&	ACAATT	& 	CGTCCA	&	CTTGTA	&	CTACAC	&	GACAGA	&	AACTAT	&	-30.58	\\
B* side 1	&	GTACGA	& 	TCACAG	&	TGAGAA	&	GAATCT	&	GGAATA	&	AGATGC	&	-30.72	\\
B* side 2	&	AATGTT	& 	GTGCAT	&	CATATC	&	ATGGAA	&	CAGGAA	&	CTATCT	&	-30.31	\\
B* side 3	&	ATAGTT	& 	TCTGTC	&	GTGTAG	&	TACAAG	&	TGGACG	&	AATTGT	&	-30.58	\\
C side 1	&	AGTTCC	& 	CGATTA	&	ATTCTG	&	ATTCAG	&	CTTGAG	&	GTAGAT	&	-30.27	\\
C side 2	&	GGATAA	& 	TCATCC	&	GGTATT	&	GGTAAT	&	ACTGAG	&	AGAGAT	&	-29.84	\\
C side 3	&	TTGGCA	& 	GACCTC	&	CCTATG	&	CTTAGG	&	TAACAG	&	TCTTCT	&	-31.1	\\
C* side 1	&	ATCTAC	& 	CTCAAG	&	CTGAAT	&	CAGAAT	&	TAATCG	&	GGAACT	&	-30.27	\\
C* side 2	&	ATCTCT	& 	CTCAGT	&	ATTACC	&	AATACC	&	GGATGA	&	TTATCC	&	-29.84	\\
C* side 3	&	AGAAGA	& 	CTGTTA	&	CCTAAG	&	CATAGG	&	GAGGTC	&	TGCCAA	&	-31.1	\\
D side 1	&	CACATC	& 	CAATAG	&	ACGAAG	&	TGATTG	&	ACCTGA	&	CTAGGA	&	-31.38	\\
D side 2	&	GAGACA	& 	ATGACA	&	GACAGA	&	TACAGG	&	ACTAAC	&	AACCTA	&	-30.57	\\
D side 3	&	CTTACT	& 	GTACAT	&	AGTATC	&	AGTCAG	&	GTATGT	&	CGATGG	&	-31.16	\\
D* side 1	&	TCCTAG	& 	TCAGGT	&	CAATCA	&	CTTCGT	&	CTATTG	&	GATGTG	&	-31.38	\\
D* side 2	&	TAGGTT	& 	GTTAGT	&	CCTGTA	&	TCTGTC	&	TGTCAT	&	TGTCTC	&	-30.57	\\
D* side 3	&	CCATCG	& 	ACATAC	&	CTGACT	&	GATACT	&	ATGTAC	&	AGTAAG	&	-31.16	\\
E side 1	&	TTCTCA	& 	GCATCT	&	CTGTGA	&	TATTCC	&	TCGTAC	&	AGATTC	&	-30.72	\\
E side 2	&	GATATG	& 	AGATAG	&	ATGCAC	&	TTCCTG	&	AACATT	&	TTCCAT	&	-30.31	\\
E side 3	&	CTACAC	& 	ACAATT	&	GACAGA	&	CGTCCA	&	AACTAT	&	CTTGTA	&	-30.58	\\
E* side 1	&	GAATCT	& 	GTACGA	&	GGAATA	&	TCACAG	&	AGATGC	&	TGAGAA	&	-30.72	\\
E* side 2	&	ATGGAA	& 	AATGTT	&	CAGGAA	&	GTGCAT	&	CTATCT	&	CATATC	&	-30.31	\\
E* side 3	&	TACAAG	& 	ATAGTT	&	TGGACG	&	TCTGTC	&	AATTGT	&	GTGTAG	&	-30.58	\\
F side 1	&	ATTCAG	& 	AGTTCC	&	CTTGAG	&	CGATTA	&	GTAGAT	&	ATTCTG	&	-30.27	\\
F side 2	&	GGTAAT	& 	GGATAA	&	ACTGAG	&	TCATCC	&	AGAGAT	&	GGTATT	&	-29.84	\\
F side 3	&	CTTAGG	& 	TTGGCA	&	TAACAG	&	GACCTC	&	TCTTCT	&	CCTATG	&	-31.1	\\
F* side 1	&	CAGAAT	& 	ATCTAC	&	TAATCG	&	CTCAAG	&	GGAACT	&	CTGAAT	&	-30.27	\\
F* side 2	&	AATACC	& 	ATCTCT	&	GGATGA	&	CTCAGT	&	TTATCC	&	ATTACC	&	-29.84	\\
F* side 3	&	CATAGG	& 	AGAAGA	&	GAGGTC	&	CTGTTA	&	TGCCAA	&	CCTAAG	&	-31.1	\\
G side 1	&	ACGAAG	& 	CTAGGA	&	ACCTGA	&	CAATAG	&	CACATC	&	TGATTG	&	-31.38	\\
G side 2	&	GACAGA	& 	AACCTA	&	ACTAAC	&	ATGACA	&	GAGACA	&	TACAGG	&	-30.57	\\
G side 3	&	AGTATC	& 	CGATGG	&	GTATGT	&	GTACAT	&	CTTACT	&	AGTCAG	&	-31.16	\\
G* side 1	&	CAATCA	& 	GATGTG	&	CTATTG	&	TCAGGT	&	TCCTAG	&	CTTCGT	&	-31.38	\\
G* side 2	&	CCTGTA	& 	TGTCTC	&	TGTCAT	&	GTTAGT	&	TAGGTT	&	TCTGTC	&	-30.57	\\
G* side 3	&	CTGACT	& 	AGTAAG	&	ATGTAC	&	ACATAC	&	CCATCG	&	GATACT	&	-31.16	\\
H side 1	&	CTGTGA	&	AGATTC	&	TCGTAC	&	GCATCT	&	TTCTCA	&	TATTCC	&	-30.72	\\
H side 2	&	ATGCAC	&	TTCCAT	&	AACATT	&	AGATAG	&	GATATG	&	TTCCTG	&	-30.31	\\
H side 3	&	GACAGA	&	CTTGTA	&	AACTAT	&	ACAATT	&	CTACAC	&	CGTCCA	&	-30.58	\\
H* side 1	&	GGAATA	&	TGAGAA	&	AGATGC	&	GTACGA	&	GAATCT	&	TCACAG	&	-30.72	\\
H* side 2	&	CAGGAA	&	CATATC	&	CTATCT	&	AATGTT	&	ATGGAA	&	GTGCAT	&	-30.31	\\
H* side 3	&	TGGACG	&	GTGTAG	&	AATTGT	&	ATAGTT	&	TACAAG	&	TCTGTC	&	-30.58	\\
I side 1	&	CTTGAG	&	ATTCTG	&	GTAGAT	&	AGTTCC	&	ATTCAG	&	CGATTA	&	-29.84	\\
I side 2	&	ACTGAG	&	GGTATT	&	AGAGAT	&	GGATAA	&	GGTAAT	&	TCATCC	&	-31.1	\\
I side 3	&	TAACAG	&	CCTATG	&	TCTTCT	&	TTGGCA	&	CTTAGG	&	GACCTC	&	-31.42	\\
I* side 1	&	TAATCG	&	CTGAAT	&	GGAACT	&	ATCTAC	&	CAGAAT	&	CTCAAG	&	-29.84	\\
I* side 2	&	GGATGA	&	ATTACC	&	TTATCC	&	ATCTCT	&	AATACC	&	CTCAGT	&	-31.1	\\
I* side 3	&	GAGGTC	&	CCTAAG	&	TGCCAA	&	AGAAGA	&	CATAGG	&	CTGTTA	&	-31.42	\\

%% file: IsotropicPatternInteractions-6-0.tex
Number of subunits	&	Subunit ID	&	Side 1 strands	&	Side 2 strands	&	Side 3 strands	&	&	Number of subunits	&	Subunit ID	&	Side 1 strands	&	Side 2 strands	&	Side 3 strands	\\
\hline
\endfirsthead

Number of subunits	&	Subunit ID	&	Side 1 strands	&	Side 2 strands	&	Side 3 strands	&	&	Number of subunits	&	Subunit ID	&	Side 1 strands	&	Side 2 strands	&	Side 3 strands	\\
\hline
\endhead

\endfoot
\endlastfoot
N=1	&	1	&	sA	&	sA	&	sA	&	&	N=14	&	1	&	A	&	A	&	A	\\
N=2	&	1	&	sA	&	A	&	A	&	&		&	2	&	B	&	B	&	A*	\\
	&	2	&	A	&	A*	&	sA	&	&		&	3	&	C	&	A*	&	B	\\
N=4	&	1	&	sA	&	A	&	A	&	&		&	4	&	B*	&	C	&	C	\\
	&	2	&	sB	&	B	&	A*	&	&		&	5	&	C*	&	D	&	D	\\
	&	3	&	A	&	A*	&	sA	&	&		&	6	&	A*	&	B*	&	E	\\
	&	4	&	A*	&	B*	&	sB	&	&		&	7	&	D	&	E	&	C*	\\
N=7	&	1	&	A	&	A	&	sA	&	&		&	8	&	D*	&	C*	&	D*	\\
	&	2	&	A*	&	B	&	A	&	&		&	9	&	E	&	D*	&	sA	\\
	&	3	&	B	&	A*	&	A*	&	&		&	10	&	F	&	F	&	E*	\\
	&	4	&	C	&	B*	&	B	&	&		&	11	&	E*	&	G	&	F	\\
	&	5	&	B*	&	C	&	B*	&	&		&	12	&	F*	&	E*	&	F*	\\
	&	6	&	C*	&	sA	&	C	&	&		&	13	&	G	&	F*	&	sB	\\
	&	7	&	sA	&	C*	&	C*	&	&		&	14	&	G*	&	G*	&	B*	\\

%% file: IsotropicPatternInteractions-10-0.tex
Number of subunits	&	Subunit ID	&	Side 1 strands	&	Side 2 strands	&	Side 3 strands	&	&	Number of subunits	&	Subunit ID	&	Side 1 strands	&	Side 2 strands	&	Side 3 strands	\\
\hline
\endfirsthead

Number of subunits	&	Subunit ID	&	Side 1 strands	&	Side 2 strands	&	Side 3 strands	&	&	Number of subunits	&	Subunit ID	&	Side 1 strands	&	Side 2 strands	&	Side 3 strands	\\
\hline
\endhead

\endfoot
\endlastfoot
N=1	&	1	&	sA	&	sA	&	sA	&	&	N=16	&	1	&	A	&	A	&	A	\\
N=3	&	1	&	sA	&	A	&	A	&	&		&	2	&	B	&	B	&	A*	\\
	&	2	&	A	&	A*	&	sA	&	&		&	3	&	B*	&	C	&	B	\\
	&	3	&	A*	&	sA	&	A*	&	&		&	4	&	C*	&	D	&	C	\\
N=4	&	1	&	A	&	A	&	A	&	&		&	5	&	A*	&	B*	&	D	\\
	&	2	&	A*	&	B	&	B	&	&		&	6	&	D*	&	E	&	B*	\\
	&	3	&	B	&	A*	&	B*	&	&		&	7	&	E*	&	C*	&	C*	\\
	&	4	&	B*	&	B*	&	A*	&	&		&	8	&	F	&	F	&	D*	\\
N=9	&	1	&	sA	&	A	&	A	&	&		&	9	&	D*	&	G	&	E	\\
	&	2	&	B	&	A*	&	B	&	&		&	10	&	E*	&	D*	&	F	\\
	&	3	&	C	&	B	&	A*	&	&		&	11	&	F*	&	E*	&	G	\\
	&	4	&	B*	&	sA	&	C	&	&		&	12	&	G	&	H	&	E*	\\
	&	5	&	A	&	D	&	B*	&	&		&	13	&	G*	&	G*	&	F*	\\
	&	6	&	C*	&	C	&	sA	&	&		&	14	&	H*	&	F*	&	G*	\\
	&	7	&	A*	&	B*	&	D	&	&		&	15	&	C*	&	A*	&	H	\\
	&	8	&	D	&	C*	&	C*	&	&		&	16	&	H*	&	H*	&	H*	\\
	&	9	&	D*	&	D*	&	D*	&	&		&       &       &       &       \\

%% file: LengthConfinedTublets.tex
Linear tiling monomers	&		&		&		&	&	Capping Monomers	&		&		&		\\ \hline
Subunit ID	&	Side 1 strands	&	Side 2 strands	&	Side 3 strands	&	&	Subunit ID	&	Side 1 strands	&	Side 2 strands	&	Side 3 strands	\\
\hline
\endfirsthead

Linear tiling monomers	&		&		&		&	&	Capping Monomers	&		&		&		\\ \hline
Subunit ID	&	Side 1 strands	&	Side 2 strands	&	Side 3 strands	&	&	Subunit ID	&	Side 1 strands	&	Side 2 strands	&	Side 3 strands	\\
\hline
\endhead

\endfoot
\endlastfoot
1	&	sA	&	sA	&	A	&	&	1-cap	&	sA	&	sA	&	Passive	\\
2	&	A	&	A	&	A*	&	&		&		&		&		\\
3	&	A*	&	A*	&	B	&	&	3-cap	&	A*	&	A*	&	Passive	\\
4	&	B	&	B	&	B*	&	&		&		&		&		\\
5	&	B*	&	B*	&	C	&	&		&		&		&		\\
6	&	C	&	C	&	C*	&	&		&		&		&		\\
7	&	C*	&	C*	&	D	&	&	7-cap	&	C*	&	C*	&	Passive	\\
8	&	D	&	D	&	D*	&	&		&		&		&		\\
9	&	D*	&	D*	&	E	&	&		&		&		&		\\
10	&	E	&	E	&	E*	&	&		&		&		&		\\
11	&	E*	&	E*	&	F	&	&		&		&		&		\\
12	&	F	&	F	&	F*	&	&		&		&		&		\\
13	&	F*	&	F*	&	G	&	&		&		&		&		\\
14	&	G	&	G	&	G*	&	&		&		&		&		\\
15	&	G*	&	G*	&	H	&	&		&		&		&		\\
16	&	H	&	H	&	H*	&	&		&		&		&		\\
17	&	H*	&	H*	&	I	&	&		&		&		&		\\
18	&	I	&	I	&	I*	&	&	19-cap	&	I*	&	I*	&	Passive	\\
Even periodic & sB & sB & X* \\
Odd periodic & X* & X* & sA \\